\documentclass{aastex62}

\usepackage[utf8]{inputenc}
\usepackage[english]{babel}
\usepackage{amsmath}
\usepackage{amsfonts}
\usepackage{amssymb}
\usepackage{graphicx}
\usepackage{gensymb}
\usepackage{tabularx}
\usepackage{array}
\usepackage{natbib}
\setcitestyle{round}
\usepackage{hyperref}
\newcolumntype{Y}{>{\centering\arraybackslash}X}

\providecommand{\keywords}[1]
{
  \small	
  \textbf{\textit{Keywords---}} #1
}

\RequirePackage{natbib}
\renewcommand\cite{\citep}

\begin{document}

\title{\Large A race against the clock: \\Constraining the timing of cometary bombardment relative to Earth's growth}

\author[0000-0001-5985-2863]{Sarah Joiret}
\affiliation{Laboratoire dAstrophysique de Bordeaux, Univ. Bordeaux, CNRS \\ 
B18N, allée Geoffroy Saint-Hilaire \\
33615 Pessac, France}
\author{Sean N. Raymond}
\affiliation{Laboratoire dAstrophysique de Bordeaux, Univ. Bordeaux, CNRS \\ 
B18N, allée Geoffroy Saint-Hilaire \\
33615 Pessac, France}
\author{Guillaume Avice}
\affiliation{Université Paris Cité, Institut de physique du globe de Paris, CNRS \\
Paris, F-75005, France}
\author{Matthew S. Clement}
\affiliation{John Hopkins APL, 11100 Johns Hopkins Rd \\ 
Laurel, MD 20723, USA}
\author{Rogerio Deienno}
\affiliation{Southwest Research Institute, 1050 Walnut St. Suite 300 \\
Boulder, CO 80302, USA}
\author{David Nesvorn\'y}
\affiliation{Southwest Research Institute, 1050 Walnut St. Suite 300 \\
Boulder, CO 80302, USA}

    \begin{abstract}
Comets are considered a potential source of inner solar system volatiles, but the timing of this delivery relative to that of Earth's accretion is still poorly understood. Measurements of xenon isotopes in comet 67P/Churyumov-Gerasimenko revealed that comets partly contributed to the Earth's atmosphere. However, there is no conclusive evidence of a significant cometary component in the Earth's mantle. These geochemical constraints would favour a contribution of comets mainly occurring \textit{after} the last stages of Earth's formation. Here, we evaluate whether dynamical simulations satisfy these constraints in the context of an \textit{Early Instability} model. We perform N-body simulations of the solar system's dynamic evolution considering an outer disk of 10000 comets, different initial conditions for the inner solar system and assuming that the giant planet instability happened within $\simeq$10 million years of the start of planet formation. Out of 30 simulations, 13 meet a list of criteria for success. We calculate the probability of collision between comets and Earth analogs component embryos through time and estimate the total cometary mass accreted in Earth analogs as a function of time. We determine that the cumulative cometary mass accreted by the Earth analogs during the first 100 million years ranges between 4 × $10^{21}$ and 5 × $10^{24}$ g. While our results are in excellent agreement with geochemical constraints, we also demonstrate that the contribution of comets on Earth might have been delayed with respect to the timing of the instability, due to a stochastic component of the bombardment. More importantly, we show that it is possible that enough cometary mass has been brought to Earth after it had finished forming so that the xenon constraint is not necessarily in conflict with an \textit{Early Instability} scenario. Indeed, 2 of the 13 successful simulations provide enough comet delivery after a last giant impact happening later than 30 Myr. However, it appears very likely that a few comets were delivered to Earth early in its accretion history, thus contributing to the mantle's budget. Finally, we compare the delivery of cometary material on Earth to Venus and Mars. These results emphasize the stochastic nature of the cometary bombardment in the inner solar system. 
    \end{abstract}
    \vspace{4mm}

\section{Introduction}
\label{Intro}
Timing is everything, especially when it comes to cometary bombardment on Earth. A bombardment of comets is thought to have been triggered by the giant planet instability \cite{Gomes} early in the history of the solar system. The timing of this instability has often been constrained to the first $\simeq$ 100 Myr after gas disk dispersal \cite{Morby2018, Nesvorny2018, Boehnke2016, Zellner, Mojzsis2019}, but recent models seem to favour an extremely early instability \cite{Clement2018, Liu2022}, probably during the first 10 Myr of the solar system. On the other hand, both geochemical constraints and N-body simulations indicate that accretion of the Earth ended between $\simeq$ 30 to 100 Myr after the condensation of Ca–Al-rich inclusions (CAIs), the first solids in the solar system \cite{Kleine, Rudge, Jacobson}. One might naively interpret this to imply that the cometary bombardment happened before Earth had finished forming. 

However, isotopic signatures of xenon in the mantle and in the atmosphere of the Earth challenge this chronology. While recent studies tend to demonstrate that primordial mantle xenon is close to the chondritic value \cite{Holland2009, Peron, Broadley}, primordial atmosphere xenon shows both a chondritic and a cometary component \cite{Marty}. Measurements obtained with the Rosetta spacecraft on gases emitted by comet 67P/Churyumov-Gerasimenko revealed a peculiar isotopic composition of cometary xenon with selective depletion in neutron-rich isotopes ($^{134}Xe$ and $^{136}Xe$). These depletions are analog to those derived for U-Xe, the primordial xenon component in our early atmosphere, i.e. the atmospheric composition of xenon corrected for mass-dependent isotope fractionation (for a review on atmospheric Xe and U-Xe, see \citet{Avice2020}). It was shown that the U-Xe isotopic composition matches a mixture of $\simeq$ 22 $\%$ cometary xenon and $\simeq$ 78 $\%$ chondritic xenon \cite{Marty}. Isotopic fractionation of U-Xe has been progressive through time on the fully-formed Earth, leading to the xenon composition of our present-day atmosphere \cite{Avice2018}. Hence, comets would have contributed to the composition of the Earth's atmosphere, but not necessarily the Earth's mantle (see Section \ref{gc} for a more detailed description of the geochemical constraints on the Earth's mantle). These results point towards a cometary bombardment that mainly occurred after the main accretion phases of the Earth, which contradicts the current understanding of the solar system timeline.  

In this paper, we propose to address this apparent paradox by constraining the range of plausible timings for cometary bombardments on the early Earth. We approach this issue using N-body simulations of planetary formation in the context of existing solar system dynamical models, combined with calculations of the collision probabilities between comets and Earth's embryos. 

Our simulations couple the evolution of the giant planets during their dynamical instability with the accretion of terrestrial planets from planetary embryos and planetesimals. They aim to successfully reproduce observational characteristics of our solar system, which mainly include the masses and orbits of planets and populations of small bodies. In particular, the orbital eccentricities and inclinations of the giant planets on one hand, and the small mass of Mars on the other hand, represent key constraints for the models \cite{Raymond2009, NM12}.

The interaction of Jupiter-mass planets (or smaller ones) with gas or planetesimals has the effect of damping their eccentricities and inclinations, and causes migration \cite{Papaloizou, Morby2009a, Kley2012}. Planets may form a resonant chain as a result of these migrations. In the case of the solar system, it was shown that the giant planets should have been in a multi-resonant configuration at the moment of gas disk dispersal \cite{Morby2007, Pierens2008}. However, the giants of our solar system are on moderately eccentric and inclined orbits; and are not locked in mutual mean motion resonances whatsoever. Similarly, the complex orbital structure of the Kuiper belt seems to imply a complex formation history. These realizations led to the hypothesis of an excitation mechanism between the giant planets which would have disrupted the pristine outer disk of planetesimals \cite{Tsiganis, Gomes, Morby2007, Levison2011}. This so-called \textit{Nice model} has prevailed over the last decades, especially considering its success at replicating a wide range of observable properties, such as the orbital architecture of the asteroid belt and Kuiper belt, the population of Jupiter's trojans asteroids, the capture of irregular moons and the obliquities of the giant planets \cite{Nesvorny2018}. Among the major updates of this model are the initial conditions, which now involve five or six giant planets instead of four \cite{Nesvorny2011, Batygin, NM12}. Stabilization of the system is then brought back following the ejection of one or two extra ice giants. 

Another crucial modification involves the timing of the instability. It was long believed that the dynamical instability, and thus the destabilization of the outer disk of planetesimals, should coincide with a cataclysmic bombardment of comets and asteroids that would have happened about 700 Myr after the formation of the solar system \cite{Gomes}. This event, known as the \textit{Late Heavy Bombardment} (LHB), has been recorded in the cratering history of the Moon and has largely been studied through the Apollo and Luna sample return missions \cite{Tera}. However, recent studies \cite{Boehnke2016, Zellner, Hartmann2019} no longer treat the LHB as a cataclysmic event and, more importantly, show that impact ages of lunar samples might reflect a sampling bias. In addition, an instability earlier than 20 - 100 Myr after CAIs can be inferred from new observations and measurements, such as the differences in highly-siderophile elements between the mantle of the Earth and the Moon \cite{Morby2018}, the existence of a binary Trojan of Jupiter \cite{Nesvorny2018} or the reset ages of minerals in certain meteorites \cite{Mojzsis2019}. For these reasons, the LHB is not considered as a firm constraint for the timing of the \textit{Nice model} anymore \cite{Morby2018}. 

Dynamically-speaking, it has also been demonstrated that the survival of the terrestrial planets is very unlikely considering a late instability \cite{Kaib2016}. An early instability is preferred over a delayed one for both the Solar System \cite{Deienno2017, Ribeiro2020} and the general case of giant planets and outer planetesimal disks \cite{Raymond2010}. In agreement with this scenario, \citet{Liu2022} show that the dynamical instability may have been triggered by the dispersal of the gas disk. The \textit{Early Instability} model makes the assumption that the giant planet instability happened early enough to directly influence terrestrial planet formation. In particular, the small mass of Mars is best replicated when the instability happens 1-10 Myr after the dispersal of the primordial gas disk \cite{Clement2018}. 

The small mass of Mars is a strong observational constraint that needs to be verified in the simulations' outcomes \cite{Raymond2009}. The classical model, which assumes a smooth distribution of planetesimals as initial conditions for the inner solar system, has consistently failed to reproduce Mars analogs \cite{Wetherill1991, Chambers1998, OBrien2006, Raymond2006, Raymond2009, Morishima2010, Fischer2014, Kaib2015}. However, several models - that are not mutually exclusive - have been proposed to overcome this problem. As stated above, the \textit{Early Instability} model is very successful at depleting Mars' feeding zone and the asteroid belt, while having little effect on the growth of Venus and Earth \cite{Clement2018, Clement2019a}. Alternatively, it is possible that there was initially less material in that area of the solar system leading to a smaller Mars and depleted asteroid belt \cite{Hansen}.  A physical justification for this type of initial conditions can be provided by ALMA observations revealing how dust in protoplanetary disks is concentrated into rings \cite{Huang}. A number of dynamical models explain how these rings may have been shaped in the early solar system. In the \textit{Grand Tack} model for example, an early temporary intrusion of Jupiter and Saturn in the inner solar system could truncate the disk of terrestrial embryos at $\simeq$ 1 AU and confine it to a ring \cite{Walsh2011}. More recently, the \textit{annulus} model has proposed that our solar system formed from rings of planetesimals \cite{Drazkowska2017, Morby2022, Izidoro2022}. These rings would have been located near the silicate sublimation line, water snowline and CO snowline and would respectively explain the formation of terrestrial planets, giant planets and primordial outer disk of planetesimals. \citet{Lykawka2023} demonstrate that chaotic excitation of disk objects triggered by a near-resonant Jupiter-Saturn configuration may also result in a narrow disk from which the terrestrial planets and the asteroid belt can form.

In this study, our simulations start with a ring of terrestrial planetesimals in the inner solar system; the \textit{Early instability} model is then included by forcing the giant planets into a dynamical instability very early in their history \cite{Nesvorny2021}. 

Analytical derivations and numerical modeling of rocky planet formation predict a runaway growth, in which a starting population of small planetesimals evolves in a bimodal mass distribution comprising both Moon to Mars-sized embryos and smaller-sized planetesimals \cite{Greenberg1978, Wetherill1993, Kokubo1996}. This phase is self-limiting and is followed by the \textit{oligarchic growth}, during which small bodies can only participate in the growth of the embryos \cite{Kokubo1998}. \citet{Walsh2019} have shown that by the end of the gas disk lifetime (considering a gas disk density decreasing exponentially with a 2 Myr timescale) the population comprises nearly 90$\%$ planetary embryos by mass around 1 AU . This mass distribution between embryos and planetesimals is taken as the starting configuration of our simulations (see section \ref{IC} for simulation details). 

The final stage of terrestrial planet formation occurred through giant impacts among the large embryos \cite{Wetherill1985, Chambers1998, Agnor1999, Asphaug2006}. The last giant impact on Earth would be the Moon-forming impact and would have led to the final phases of Earth's differentiation \cite{Canup2001, Cuk2012, Canup2022}. The timing of this event can be derived from the Hf-W chronometry and dynamical simulations and is estimated to range between 30 and 100 Myr after CAIs, thus corresponding to the accretion timing of the Earth \cite{Kleine, Rudge, Jacobson}. 

During the last few years, an alternative school of thought has emerged to explain the growth of rocky embryos. It was proposed that accretion of inward-drifting \textit{pebbles} from the outer solar system could be extremely efficient in the presence of a gaseous disk and result in a very rapid formation of the terrestrial planets \cite{Lambrechts2019, Johansen2021}. In this scenario, Earth would have finished forming by the time of gas disk dispersal, leaving plenty of time for the comets to contribute to Earth's atmosphere without contributing to its mantle. Hence, the xenon isotopic dichotomy between the mantle and the atmosphere of the Earth could be explained more easily. However, recent studies have proven the contribution of outer solar system material in rocky planets to be small \cite{Burkhardt2021, Mah2022}. And even if Earth had formed in a few millions years, it would still require one late giant impact to form the Moon.  Thus, it remains unclear whether the pebble accretion model is consistent with the observed isotopic composition of inner solar system bodies.

\section{Material and methods}
\subsection{Geochemical constraints}
\label{gc}
\begin{figure}[h!]
\centering
  \includegraphics[width=8cm]{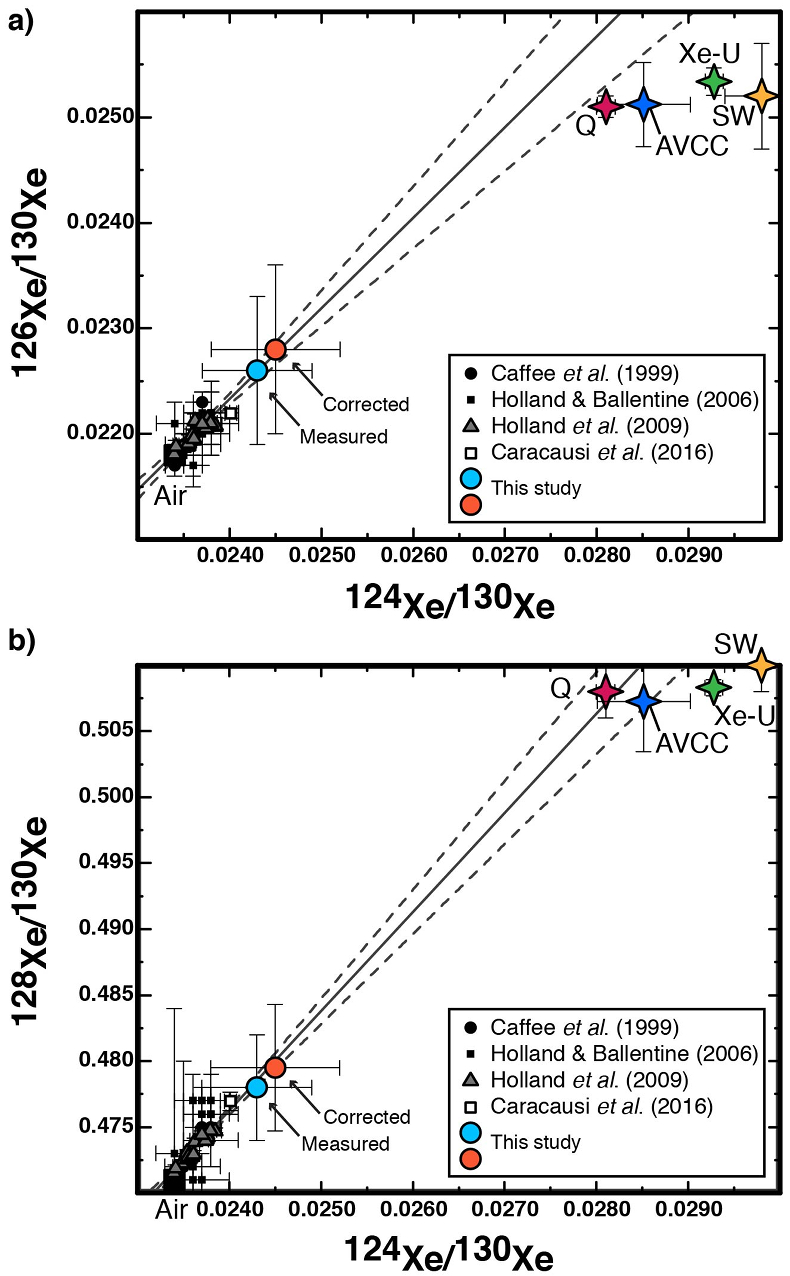}
  \caption{\small Light xenon constraints on the mantle of the Earth, taken from \citet{Peron}. The blue dot represents measured data and the orange dot represents the corrected data for atmospheric contamination. The dashed lines are extrapolated mantle-air mixing lines that indicate the uncertainties on the mantle measurements with a 95\% confidence interval. These data suggest a chondritic origin (Q-phase or AVCC) for mantle xenon. (\url{https://creativecommons.org/licenses/by-nc-nd/4.0/})}
  \label{Peron}
\end{figure}

The xenon isotopic dichotomy between the mantle and the atmosphere of the Earth should be qualified by some considerations. The chondritic origin of the mantle has been determined by comparing the mantle-air Xe mixing lines to different cosmochemical endmembers. The uncertainties on the mantle Xe measurements give rise to an error envelope between the extrapolated minimum and maximum mantle-air mixing lines (the two dashed lines in panels a and b of Figure \ref{Peron}, taken from \citet{Peron}). If a cosmochemical endmember falls into that error envelope, it means that the mantle composition could be a mixture between air (Xe atmospheric value) and that endmember. The different endmembers shown in Figure \ref{Peron} are Q-Xe, AVCC-Xe, U-Xe and SW-Xe. Q-phase is the carbon-rich residue of a meteorite after acid digestion, it is the main carrier of heavy noble gases in chondrites. AVCC is an acronym for Average Value Carbonaceous Chondrites. U-Xe is the primordial component of Xe in the atmosphere of the Earth and it was shown to be a mixture between 22\% cometary Xe and 78\% chondritic Xe \cite{Marty}. And SW stands for solar wind, it is the solar value. 

Panel a of Figure \ref{Peron} shows that the extrapolation of mantle-air Xe mixing lines does not fit any known cosmochemical endmember within uncertainty. Panel b, on the other hand, shows that the extrapolation of mantle-air Xe mixing lines could fit Q-Xe and AVCC-Xe, while U-Xe and SW-Xe are outside the error envelope. This is how it was concluded that the Earth mantle is chondritic. However, it does not prove that there was absolutely no cometary contribution to the Earth’s mantle. For example, a mixture between 50\% Q-Xe and 50\% U-Xe would still fall within the error envelope of this mantle-air mixing line. As we know that U-Xe contains around 22\% of cometary Xe, that would mean that the Earth’s mantle could include up to $\simeq$ 10\% of cometary material. 

Moreover, the Earth's deep mantle deficit in $^{86}$Kr relative to chondritic Kr proves that a purely chondritic origin for the mantle is not completely established, as comet 67P/Churyumov-Gerasimenko is the only known planetary body that has also revealed a deficit in $^{86}$Kr to this day \cite{Peron2021}. Hence, comets would have contributed to the deep mantle's Kr budget during the earliest phase of Earth's growth.

Be that as it may, even with a 10\% cometary component in the Earth's mantle, the primordial atmosphere would still contain at least two times more cometary xenon. Therefore, the xenon isotopic dichotomy between the mantle and the atmosphere of the Earth, and its related paradox, remain valid.

\subsection{N-body simulations}
We perform N-body simulations to constrain the timing of cometary bombardment relative to accretion of Earth analogs. We start the simulations at the time of gas disk dispersal ($t_{0}$). At this stage, the giant planets are already formed and the inner solar system comprises planetesimals and Moon- to Mars-sized embryos. We also consider an outer disk of planetesimals beyond the orbits of the giant planets to account for the comets. All the dynamics of the system occur in a gas-free environment and only gravitational forces - which dominate the trajectories for particles $ > $ 100 $\mu$m \cite{Koschny2019} - are taken into account. Simulations last for 100 Myr after $t_{0}$.
\subsubsection{Collisional algorithm}
Simulations are performed using \textit{GENGA} \cite{Grimm2014}, a hybrid symplectic N-body integrator. \textit{Hybrid} means that close encounters are handled with good energy conservation; and \textit{symplectic} means that the numerical integrator conserves the Hamiltonian structure of the equations of motion. \textit{GENGA} is based on the integration scheme of \textit{Mercury} \cite{Chambers1999} but is GPU-accelerated, implying that many operations are performed in parallel. Aside from being faster than \textit{Mercury}, \textit{GENGA} also includes an interpolation option that can force a set of bodies to follow - or evolve towards - a certain orbit. This feature is used to include successful simulations of the giant planets that accurately reproduce the outer solar system \cite{Deienno2018, Clement2021b, Nesvorny2021}.
\subsubsection{Giant planet instability}
As a reference, we consider two dynamical cases with no instability. One of them includes the terrestrial embryos and planetesimals only, and the other one also comprises the four giant planets set on their current orbits at the beginning of the simulation. We refer to them as the \textit{nojov} and \textit{jovi} sets, respectively. These sets serve as control cases for the sets including an instability.  

\begin{figure}[h!]
\centering
  \includegraphics[width=8cm]{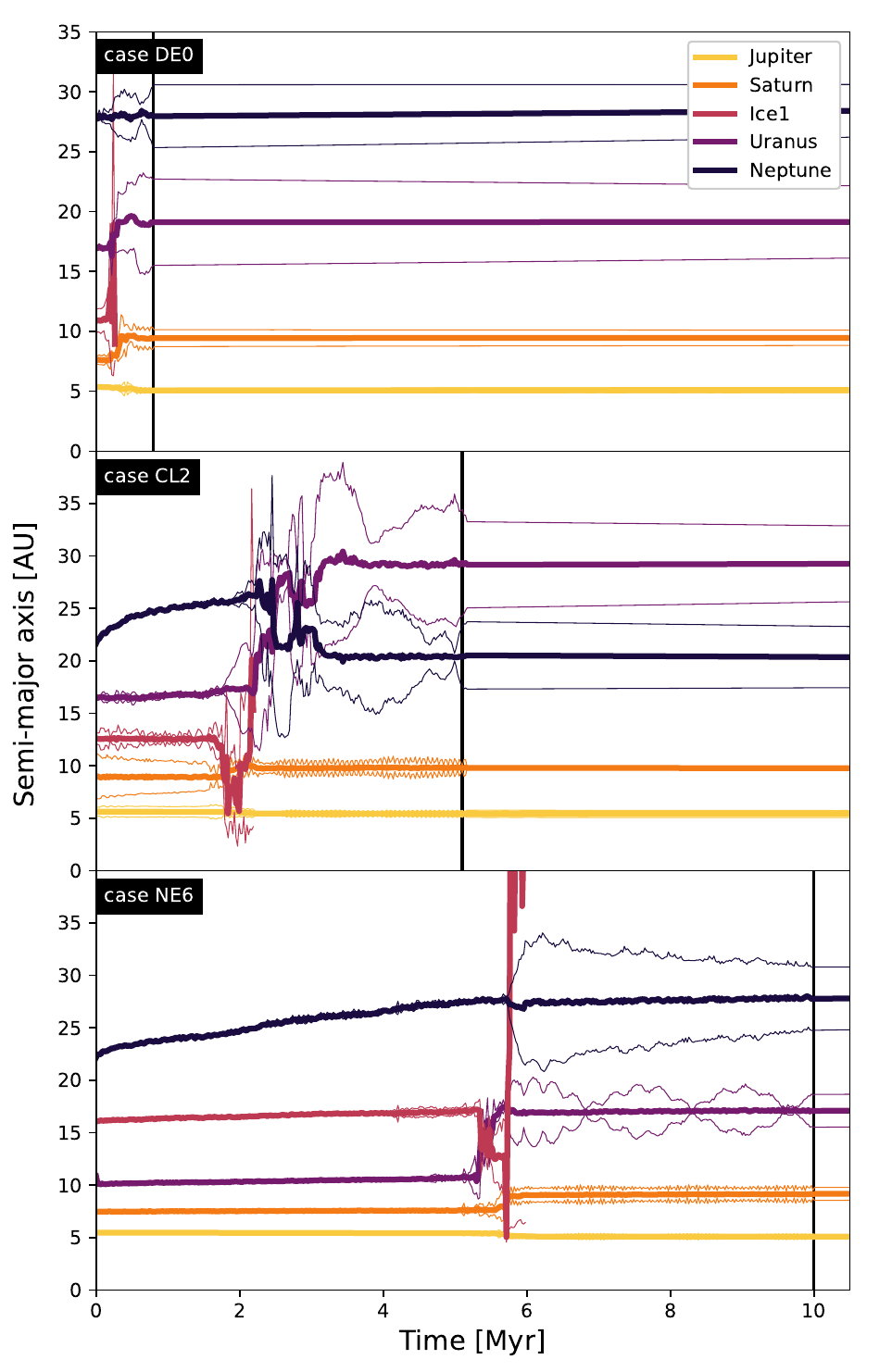}
  \caption{\small Early evolution of the giant planets in our three instability simulations. The onset of the instability happens at $t_{inst}$ $\simeq$ 0.6 Myr in the \textit{DE0} set, $t_{inst}$ $\simeq$ 2 Myr in the \textit{CL2} set and $t_{inst}$ $\simeq$ 6 Myr in the \textit{NE6} set. In each case, the extra ice giant is ejected right after $t_{inst}$ and the four remaining giants start stabilizing again. The black vertical lines indicate the starting point of the second interpolation during which the giant planets evolve smoothly towards their current orbit. We only show the dynamical evolution of the first $\simeq$ 10 Myr for a visualization purpose. In \citet{Deienno2018}, the instability actually starts at t $\simeq$ 5.2 Myr after a smooth migration of Neptune through the disk; however, we have neglected this outward migration in case \textit{DE0}, and Neptune is already at 28 AU at $t_{0}$.}
  \label{instability_cl}
\end{figure}

Regarding the instability cases,  we take simulation outputs for the giant planets from \citet{Deienno2018}, \citet{Clement2021b} and case 1 in \citet{Nesvorny2021} and refer to them as the \textit{DE0}, \textit{CL2} and \textit{NE6} sets respectively. Conducting the integration of the giant planets separately from a more comprehensive simulation ensures that the orbital evolution of the giant planets is reproduced exactly like in the successful runs. In particular, many outer solar system constraints, such as the wide radial spacing and orbital eccentricities of giant planets, must be satisfied \cite{NM12}. More importantly, it ensures that the instability is triggered at a precise timing. As we consider the \textit{Early instability} model in this study, we set $t_{inst} \simeq$ 0.6 Myr (almost directly after the gas disk dispersal) in the \textit{DE0} set, $t_{inst} \simeq$  2 Myr in the \textit{CL2} set and $t_{inst} \simeq$ 6 Myr in the \textit{NE6} set (cf. Figure \ref{instability_cl}).  

Migration models predict that the giant planets' orbits are in mutual resonances at the moment of gas disk dispersal \cite{Papaloizou, Morby2007, Pierens2008}. In particular, it was shown that Jupiter's orbital ratio with Saturn is determinant for the dynamical evolution of the inner solar system \cite{Brasser2009, Morby2010}. The 3:2 Jupiter-Saturn resonance is often considered as the only possible primordial configuration for the outer solar system \cite{Masset2001, Morby2007, Pierens2008}.  Hence, in both \citet{Deienno2018} and \citet{Nesvorny2021}, the selected simulation of the giant planets starts from an initial five-planet multi-resonant configuration with period ratios of 3:2, 3:2, 2:1, 3:2 at $t_{0}$, as it was shown to generate the most plausible dynamical evolutions \cite{NM12}. However, it was found that a 2:1 Jupiter-Saturn resonance is also possible assuming specific combinations of gas disk parameters \cite{Pierens2014}. \citet{Clement2021b} studied this primordial 2:1 MMR (mean motion resonance) with inflated eccentricities, starting with five giant planets in a 2:1, 4:3, 3:2, 3:2 chain. This configuration was shown to be highly successful at generating a final system of planets with low eccentricities and inclinations. 

In \citet{Deienno2018}, the onset of the instability actually happens 5.2 Myr after $t_{0}$ (i.e. after the initial multi-resonant configuration); but in our \textit{DE0} set, $t_{inst}$ was directly initiated to investigate the scenario where instability is triggered by the gas disk dispersal \cite{Liu2022}. Hence, in our \textit{DE0} configuration, Neptune is already at 28 AU at $t_{0}$ and its smooth outward migration through the disk before the instability has been neglected (refer to Figure 1 in \citet{Deienno2018} for a view of the entire evolution from that work). 

The mass of the additional ice giant is comparable to that of Uranus or Neptune in the \textit{DE0} and \textit{NE6} sets (i.e. $\simeq$ 15 $M_{Earth}$), but is about half their masses in the \textit{CL2} set (i.e. $\simeq$ 8 $M_{Earth}$). This extra planet is ejected out of the system during the instability. The masses of the four other giant planets are set to their current values in every case. 

Regarding the configuration of the outer disk of planetesimals, \citet{Clement2021b} and \citet{Nesvorny2021} considered a ring beyond the orbit of Neptune, with outer edge at 30 AU, surface density profile that falls off as $\sum(r) = r^{-1}$, and total mass of 35 $M_{Earth}$; whereas a total mass of 30 $M_{Earth}$ was studied in case 1 of \citet{Deienno2018}. In all these three separate integrations, the outer disk already interacted gravitationally with the giant planets, influencing their orbits. 

By the end of the instability, the four remaining planets have stabilized and we interpolate their orbits once again to make them smoothly evolve towards their current orbit. This interpolation lasts until the end of the simulation, i.e. 100 Myr after $t_{0}$. 

This artificial procedure for the evolution of the giant planets guarantees that the cometary bombardement is directly induced by a dynamical instability matching the main solar system constraints.

\subsubsection{Initial conditions for the inner and outer planetesimal disks}
\label{IC}
Inner bodies are uniformly distributed in a ring extending from 0.7 to 1.2 AU. They follow a surface density profile $\sum(r) = r^{-1}$. The initial eccentricities and initial inclinations follow a Rayleigh distributions with $\sigma_{e}$ = 0.005 and $\sigma_{i}$ = 0.0025 respectively, as considered in \citet{Nesvorny2021}. 

In order to account for a faster or slower accretion timing relative to the cometary bombardment, two types of set were considered for the masses of the terrestrial embryos. The \textit{big} set explores the scenario where Mars-sized embryos dominate the inner planetesimals disk, whereas the \textit{small} set investigates embryos with a few lunar masses. As noted in Section \ref{Intro}, the inner population around 1 AU should comprise $\simeq$ 90$\%$ planetary embryos by mass and $\simeq$ 10$\%$ planetesimals. We set the total mass of the inner ring to 2.5 $M_{Earth}$, comprising 500 planetesimals and 20 or 40 embryos for the \textit{big} and \textit{small} sets respectively. With this set of initial conditions, inner planetesimals have a mass of 5 x $10^{-4}$ $M_{Earth}$, while embryos have a mass of 0.1125 $M_{Earth}$ $\simeq$ 1.047 $M_{Mars}$ in the \textit{big} set and a mass of 5.625 x $10^{-2}$ $M_{Earth}$ $\simeq$ 0.52 $M_{Mars}$ $\simeq$ 4.57 $M_{Moon}$ in the \textit{small} set. We take a bulk density of 3 $g/cm^{3}$ for embryos and 2 $g/cm^{3}$ for planetesimals \cite{Carry2012}. 

Regarding the comets, we consider an outer ring of planetesimals uniformly distributed between 21 and 30 AU. The initial eccentricities and inclinations also follow a Rayleigh distribution with $\sigma_{e}$ = 0.005 and $\sigma_{i}$ = 0.0025 respectively. A cold disk with an inner edge at 21 AU means that, at the beginning of the integrations of the \textit{DE0} and \textit{NE6} sets, the outermost planet is already inside the cold disk, even though its slow outward migration should have already excited this portion of the disk. This can lead to very early scattering of comets into the inner Solar System, even before the giant planet dynamical instability.  A more self-consistent approach would have been to model the excitation of the outer planetesimal disk before the onset of instability, as in \citet{Ribeiro2020}.  To account for this issue, in some of our analysis we divided the cometary bombardment into components before and after the instability. Nonetheless, proceeding this way, we did not find significant changes in the orders of magnitude of the cometary mass brought to the terrestrial planets.

There are 10000 comets representing a total mass of 25 $M_{Earth}$, which means that every comet has a mass of 2.5 x $10^{-3}$ $M_{Earth}$ $\simeq 1.5 \times 10^{25}$ g $\simeq$ $M_{Pluto}$. Their bulk density is set to 0.5 $g/cm^{3}$. Even though the masses of the comets are not necessarily accurate, it was shown that the primordial outer disk of planetesimals should have had some hundreds to thousands Pluto-mass bodies, and they could represent 10\% to 40\% of the total disk mass \cite{Nesvorny2016, Kaib2021}. Here, we consider a larger number of these massive Kuiper belt objects to balance the lack of smaller comets in our initial conditions.

\subsubsection{Summary of simulations and success criteria}
\label{success}
Overall, we explore two sets for the initial conditions of the inner solar system - the \textit{big} and \textit{small} sets - for each of the five dynamical scenarios - \textit{nojov}, \textit{jovi}, \textit{DE0}, \textit{CL2} and \textit{NE6}. This adds up to a total of ten sets. For each of these, five simulations were run with randomized orbital elements, within the framework detailed in section \ref{IC}. The number of simulations performed for each set is mainly limited by their duration and significant computational cost.

A simulation is successful when its outcome matches most of the solar system constraints. In the past few decades, several efforts have been made to measure this success in absolute terms. The most commonly used metrics are the angular momentum deficit (AMD, or $S_{d}$) introduced by \citet{Laskar1997} and the radial mass concentration (RMC, or $S_{c}$) introduced by \citet{Chambers2001}.

The angular momentum deficit is very useful to assess the dynamical excitation of the newly-formed rocky planets relative to a perfectly circular and coplanar system,  and is defined in its normalized form \cite{Chambers2001} as
\begin{equation}
    AMD = \frac{\sum_{j} m_{j}\sqrt{a_{j}}(1-\sqrt{1-e_{j}^{2}} \cos i_{j})}{\sum_{j} m_{j}\sqrt{a_j} }
\end{equation}
where $m_{j}$, $a_{j}$, $e_{j}$ and $i_{j}$ are the mass, semimajor axis, eccentricty and inclination of planet $j$ respectively. The AMD of the solar system's terrestrial planets is 0.0018.

On the other hand, the radial mass concentration provides information on the mass distribution of these planets and is defined as
\begin{equation}
    RMC = max(\frac{\sum_{j} m_{j}}{\sum_{j} m_{j}(\log_{10}(\frac{a}{a_{j}})^{2}})
\end{equation} 
where the maximum is taken over a. In particular, this value is very sensitive to the small Mars problem \cite{Raymond2009}. For comparison, the RMC of the solar system's terrestrial planets is 89.9 but a bigger Mars would decrease this number.

However, \citet{Nesvorny2021} and \citet{Clement2023} show the limitations of these statistics and propose new constraints on the terrestrial planets. Our metrics for success is largely inspired by \citet{Nesvorny2021} and obey the following scheme:
\begin{enumerate}
    \item The outcome of the simulation must include two (and only two) planets with mass M $ \gtrsim \frac{M_{Earth}}{2}$ between $\simeq$ 0.6 and 1.3 AU. These two planets are called the Venus and Earth analogs.
    \item Only one planet can be located between 1.3 and 2.0 AU, we call it a Mars analog. Its mass must be $\lesssim$ $\frac{M_{Earth}}{3}$. However, the absence of a Mars analog is not a constraint for the simulation to be considered a success.
    \item Only one planet can be located at a distance $<$ 0.6 AU, we call it a Mercury analog. Its mass must be $\lesssim$ $\frac{M_{Earth}}{4}$. However, the absence of a Mercury analog is not a constraint for the simulation to be considered a success.
    \item For every remaining successful simulations, we calculate the corresponding AMD and RMC and compare it to the solar system. We also calculate the collision rate of comets with Earth analogs from each of these successful simulations.
\end{enumerate}

\subsection{Collision rate calculations}
\label{coll_cal}
We use an algorithm based on the {\"O}pik/Wetherill approach \cite{Wetherill1967} including considerations from \citet{FarinellaDavis1992}. This method calculates the probability of collision between two planetary bodies in the generalized case where the orbits of both bodies are ellipses. In this section, we present the method without any mathematical demonstration. For more details, please refer to \citet{Wetherill1967}.
 
Let ($a_{e}$, $e_{e}$, $i_{e}$) and ($a_{c}$, $e_{c}$, $i_{c}$) be sets of orbital elements of an Earth analog embryo of radius R and of a comet of radius r, respectively. The reference frame is set so that the origin is located at the intersection between the orbit of the Earth analog embryo and the line of mutual nodes between the two orbit planes. X axis is the node line and the XY plane is the plane of the embryo orbit. A drawing of this configuration can be found in Figure 3 of \citet{Greenberg1982}. It is assumed that the trajectories of both the Earth analog embryo and the comet are rectilinear near their point of approach. XY' is the trajectory of the Earth analog embryo and is in the plane of the reference frame. AB is the trajectory of the comet, where A is the intersection between the comet orbit and the mutual node line, and B is a point of the comet orbit but is not on the plane of reference. 

\citet{Wetherill1967} assumes that the longitudes of ascending nodes ($\Omega_{e}$ and $\Omega_{c}$) and arguments of perihelion ($\omega_{e}$ and $\omega_{c}$) vary uniformly with time, so there is a uniform probability distribution for these parameters between 0 and 2$\pi$. The probability of collision between the two bodies will thus be averaged over all their possible collision geometries and these parameters are not required as input values.

We only consider pairs of bodies for which $q_{e} < Q_{c}$ and $Q_{e} > q_{c}$, where $q=a(1-e)$ is the perihelion and $Q=a(1+e)$ the aphelion of a given body. Otherwise, the orbits do not intersect each other and the collision probability is zero \cite{FarinellaDavis1992}.

The mutual inclination i' of the two bodies is given by 
\begin{equation}
    \cos{i'} = \cos{i_{e}}\cos{i_{c}} + \sin{i_{e}}\sin{i_{c}}\cos{\Delta\Omega}
\end{equation} where $\Delta\Omega$ is the difference between the longitudes of ascending nodes.
It is not necessary to know $\Delta\Omega$ because - as stated above - all relative orientations of the nodes are assumed to be equally probable. Thereby, we will further integrate the collision probability over $\Delta\Omega$, which is equivalent to varying all the possible mutual inclination i' between the two orbit planes. 

The distance of the Earth analog embryo from the Sun is defined as
\begin{equation}
    r = \frac{a_{e}(1-e_{e}^{2})}{e_{e}\cos{\theta_{e}}}
\end{equation}
where $\theta_{e}$ is the true anomaly of the Earth analog embryo. Again, it is not necessary to know $\theta_{e}$. As it is assumed that all values of the argument of perihelion $\omega_{e}$ are equally probable, we will further integrate the collision probability over $\omega_{e}$, or - in an equivalent manner - over the true anomaly in the orbit of the Earth analog embryo ($\theta_{e}$) at the point of closest approach between the two orbits. Here, Wetherill implicitly assumes that the probability distribution for the true anomaly $\theta_{e}$ of closest approach is also uniform from 0 to 2$\pi$. This assumption can introduce error factors of the order of the orbital eccentricities \cite{Greenberg1982}. However, \citet{FarinellaDavis1992} showed that these errors can be reduced to the smaller of the two eccentricities using a suitable ordering of the eccentricities in the algorithm; and, in our case, the Earth analog embryos usually have much smaller eccentricities than the colliding comets.

We set 
\begin{equation}
    c = \frac{1}{8\pi^{2}\sin{i'}a_{e}^{2}a_{c}^{2}\sqrt{(1-e_{e}^{2})(1-e_{c}^{2})}}
\end{equation}
And we have
\begin{equation}
    \cot{\alpha_{e}} = \pm\sqrt{\frac{a_{e}^{2}e_{e}^{2}-r^{2}(\frac{a_{e}}{r}-1)^{2}}{a_{e}^{2}(1-e_{e}^{2})}}
\end{equation}
\begin{equation}
    \cot{\alpha_{c}} = \pm\sqrt{\frac{a_{c}^{2}e_{c}^{2}-r^{2}(\frac{a_{c}}{r}-1)^{2}}{a_{c}^{2}(1-e_{c}^{2})}}
\end{equation}

where $\alpha_{e}$ is the angle between the X axis (or mutual node line) and XY' (the trajectory of the Earth analog embryo), and $\alpha_{c}$ is the angle between the X axis and AB (the trajectory of the comet). 

The relative velocity U between the two bodies can be obtained from
\begin{multline}
    U_{+}^{2} = \frac{GM}{r}\bigg[4 - \frac{r}{a_{e}}-\frac{r}{a_{c}} - 2\big[\frac{a_{e}a_{c}}{r^{2}}(1-e_{e}^{2})(1-e_{c}^{2})\big]^{\frac{1}{2}}\\
    (\cos{i'}+|\cot{\alpha_{e}}||\cot{\alpha_{c}}|)\bigg]
\end{multline}
and
\begin{multline}
    U_{-}^{2} = \frac{GM}{r}\bigg[4 - \frac{r}{a_{e}}-\frac{r}{a_{c}} - 2\big[\frac{a_{e}a_{c}}{r^{2}}(1-e_{e}^{2})(1-e_{c}^{2})\big]^{\frac{1}{2}}\\
    (\cos{i'}-|\cot{\alpha_{e}}||\cot{\alpha_{c}}|)\bigg]
\end{multline}
where we account for the fact that, for a given value of r, $\cot{\alpha_{e}}\cot{\alpha_{c}}$ can take two values, depending on whether $\cot{\alpha_{e}}$ and $\cot{\alpha_{c}}$ have the same or opposite signs. GM is the heliocentric gravitational constant and equals to $1.3202 \times 10^{26}$ $km^{3}kg^{-1}yr^{-2}$. Thus, we have 
\begin{equation}
    U = \frac{|U_{+}| + |U_{-}|}{2}
\end{equation}
expressed in km/year. It must be integrated over all possible true anomalies of the Earth analog embryo $\theta_{e}$, and over all possible relative longitudes of ascending node of the two bodies $\Delta\Omega$. We can divide it by $3.154\times 10^{7}$ to have it expressed in km/s.

The intrinsic collision probability per unit of time (i.e. per year in our case) will be
\begin{equation}
    P_{i} = \frac{(|U_{+}| + |U_{-}|) \,r  \,c}{|\cot{\alpha_{c}}|}
\end{equation}
and must also be integrated over all $\theta_{e}$, and $\Delta\Omega$ to obtain the total intrinsic collision probability per unit of time $P_{tot}$. The collision probability is called \textit{intrinsic} because it only depends on the two orbits involved but is independent on the size of the corresponding bodies. In other words, it defines the collision probability per unit of time between two bodies when the sum of their respective radii equals 1 km. 

Hence, the number of collisions between an Earth embryo and a comet occuring during a certain time interval $\Delta$t can be expressed as
\begin{equation}
    n_{coll} = P_{tot}(R+r)^{2} f_{grav} \Delta t
\end{equation}
where R and r are the radii in km of the Earth analog embryo and the comet respectively, and $f_{grav}$ (dimensionless) is the gravitational focus of the Earth analog embryo given by 
\begin{equation}
    f_{grav} = 1 + \frac{v_{esc,e}^{2}}{U^{2}}
\end{equation}
with $v_{esc,e} = \sqrt{\frac{2GM_{e}}{R}}$ the escape velocity of the Earth analog embryo.

The cometary mass hitting an Earth analog constituent embryo during a certain time interval can be obtained by multiplying the number of collisions $n_{coll}$ by $1.5 \times 10^{25}$ g, the mass of one comet in our simulations. 

For each timestep of a simulation (from 0 to 100 Myr) and for each embryo or planetesimal that formed the Earth analog, we calculate the probability of collision with each of the 10000 comets. By adding together the contribution of all 10000 comets in a given embryo/planetesimal, we can derive the cometary mass accreted by each embryo/planetesimal as a function of time. Then, by adding together the cometary contributions in all of the Earth analog embryos and planetesimals - which eventually bash into each other to form the Earth analog -, we obtain the total cometary mass accreted in the Earth analog as a function of time. For the sake of simplicity, in what follows, we will use the term 'Earth constituent embryos' even when we refer to Earth constituent planetesimals. 

\section{Results}
\subsection{N-body simulations: diversity of outcomes}
\label{outcomes}
The precise values for the masses, semi-major axes, AMD and RMC of the terrestrial planet analogs from each successful simulation at t = 100 Myr can be found in Table \ref{table}.

\begin{table*}[t!]
\begin{tabularx}{\linewidth}{lYYYYYYYY}
\hline
     & AMD & RMC & $M_{Earth}$ ($M_{Earth}$) & $a_{Earth}$ (AU) & $M_{Venus}$ ($M_{Earth}$)& $a_{Venus}$ (AU) & $M_{Mars}$ ($M_{Earth}$)& $M_{Mercury}$ ($M_{Earth}$)\\
 \textbf{solar system} & \textbf{0.0018} & \textbf{89.9} & \textbf{1} & \textbf{1} & \textbf{0.815} & \textbf{0.723} & \textbf{0.107} & \textbf{0.055}\\ \hline
 \textit{nojov}/big/2 & 0.0070 & 45.3 & 0.966 & 1.096 & 0.857 & 0.727 & 0.237 & 0.243 \\ 
 \textit{nojov}/big/5 & 0.0060 & 53.0 & 0.714 & 1.267 & 1.329 & 0.830 & 0.117 & 0.254 \\ 
 \textit{nojov}/small/1 & 0.0263 & 27.2 & 1.070 & 1.161 & 1.176 & 0.652 & / & / \\ 
 \textit{nojov}/small/2 & 0.0041 & 42.0 & 1.137 & 1.156 & 1.109 & 0.653 & / & / \\
 \textit{nojov}/small/4 & 0.0043 & 50.4 & 1.200 & 1.257 & 0.916 & 0.808 & / & 0.253 \\\hline
 \textit{jovi}/small/3 & 0.0014 & 113.3 & 1.224 & 0.865 & 0.764 & 0.667 & 0.181 & / \\\hline
 \textit{DE0}/big/1 & 0.0006 & 95.1 & 1.189 & 1.045 & 0.867 & 0.648 & / & / \\ 
 \textit{DE0}/big/5 & 0.0025 & 108.3 & 0.464 & 1.243 & 1.697 & 0.725 & / & / \\ 
 \textit{DE0}/small/2 & 0.0043 & 84.9 & 1.015 & 1.047 & 1.161 & 0.662 & 0.059 & / \\
 \textit{DE0}/small/3 & 0.0060 & 51.9 & 1.036 & 0.982 & 0.980 & 0.619 & 0.287 & / \\
 \textit{DE0}/small/5 & 0.0174 & 66.7 & 0.534 & 1.094 & 1.402 & 0.691 & 0.295 & / \\ \hline
 \textit{CL2}/big/1 & 0.0140 & 87.7 & 0.474 & 1.208 & 1.684 & 0.742 & 0.121 & / \\
 \textit{CL2}/big/2 & 0.0011 & 103.2 & 0.712 & 1.084 & 1.459 & 0.715 & 0.116 & / \\ 
 \textit{CL2}/big/3 & 0.0085 & 64.9 & 0.596 & 1.268 & 1.452 & 0.776 & 0.121 & 0.123 \\ 
 \textit{CL2}/small/5 & 0.0131 & 44.4 & 1.074 & 1.176 & 1.101 & 0.651 & 0.180 & / \\ \hline
 \textit{NE6}/big/1 & 0.0045 & 126.7 & 1.077 & 0.887 & 0.612 & 0.675 & 0.119 & / \\ 
 \textit{NE6}/big/4 & 0.0233 & 65.3 & 1.173 & 1.047 & 0.848 & 0.588 & / & / \\
 \textit{NE6}/small/1 & 0.0044 & 101.5 & 1.301 & 0.981 & 0.729 & 0.609 & / & / \\
 \textit{NE6}/small/5 & 0.0082 & 51.1 & 0.962 & 1.023 & 1.095 & 0.630 & 0.355 & / \\ \hline
\end{tabularx}

\caption{\small Summary of all successful simulations (i.e. meeting the criteria established in section \ref{success}). The name given to each simulation specifies the instability case, the initial conditions for the inner ring of planetesimals and a number from 1 to 5 (5 simulations were run for each set). The values specific to the solar system are displayed at the top for comparison.}
\label{table}
\end{table*}

Out of the ten \textit{nojov} simulations (including both \textit{big} and \textit{small} initial conditions for the inner ring of rocky embryos), five are successful. The median RMC value of all ten simulations is 45.5 and there are a lot of small planetesimals remaining in the inner part of the solar system at t = 100 Myr. When the contribution of the giant planets is added to the initial conditions (\textit{jovi} set), the median RMC value increases to 66.5. However, only one simulation out of ten is successful according to our success criteria and the median AMD value also increases (from 0.006 to 0.0155). Therefore, the presence of the giant planets have an important effect on the formation of the inner planets \cite{Kaib2016}.

\begin{figure}[h!]
  \centering
  \includegraphics[width=11cm]{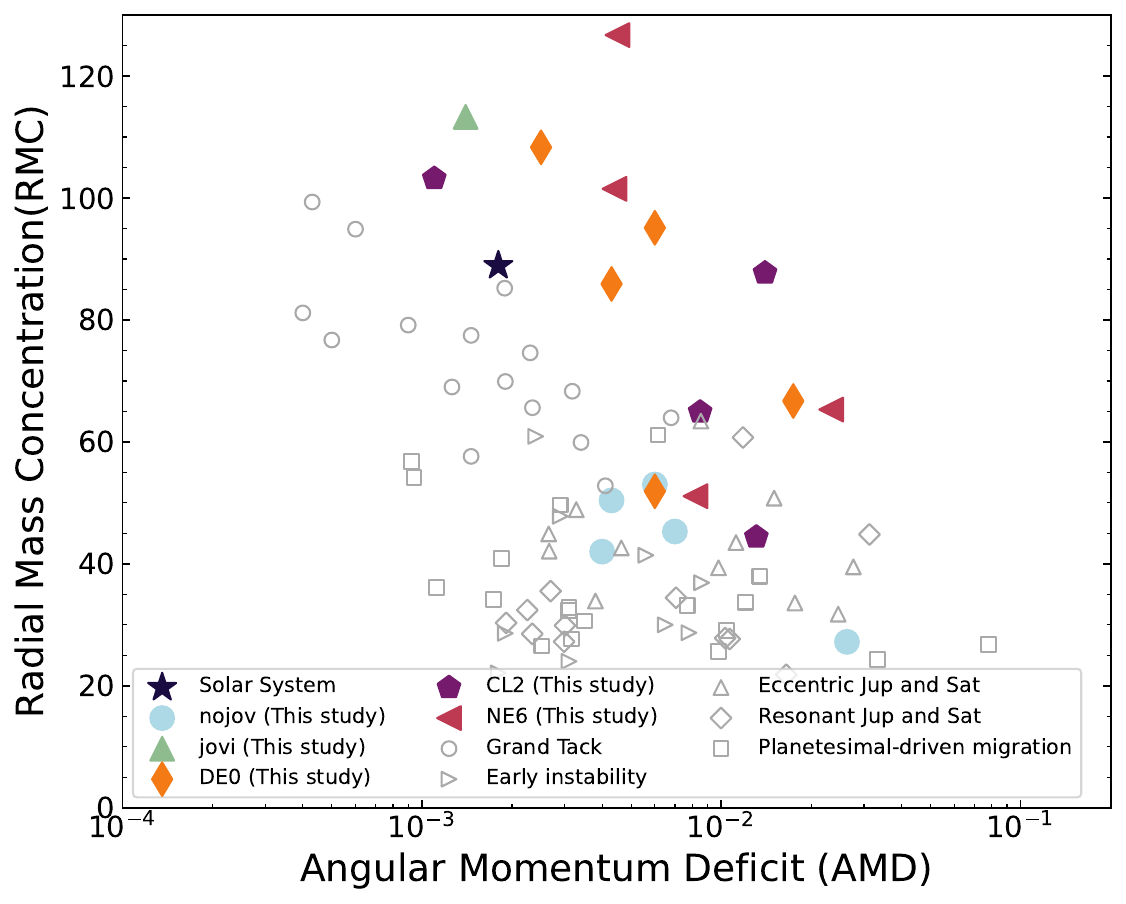}
  \caption{\small Comparison of the AMD and RMC values of our successful simulations with the solar system and other simulations from the literature. The Grand Tack simulations are from \citet{OBrien2014}. The Early Instability simulations are from \citet{Nesvorny2021}. Those with eccentric and resonant gas giants are from \citet{Raymond2009} and those including planetesimal-driven migration of the gas giants are from \citet{Lykawka2013}.}
  \label{AMD_RMC}
\end{figure}

We include an early instability among the giant planets and an outer disk of 10000 planetesimals (the \textit{DE0}, \textit{CL2} and \textit{NE6} sets) and the number of successful simulations increases, with slightly higher median RMC and smaller AMD (i.e. closer to the values of our solar system: AMD = 0.0018 and RMC = 89.9).  An overview showing the RMC and AMD values of all the successful runs is presented in Figure \ref{AMD_RMC}.

Out of the 10 simulations performed for each instability case, 5 are successful for the \textit{DE0} set, 4 for the \textit{CL2} set and 4 for the \textit{NE6} set. Our sample of simulation outputs is too small to conclude on the best timing for the instability to happen between $t_{inst}$ = 0.6, 2 or 6 Myr.  In \citet{Clement2018}, a range of timings were also investigated and simulations with  $t_{inst} <$ 1 Myr were usually less successful because the inner disk of planetesimals would not be sufficiently depleted by the end of the instability, and would thus re-spread, leading to smaller Venus and Earth analogs.  However, in this study, we consider less mass in the inner ring of planetesimals to begin with, so this is not an issue.

In the \textit{NE6} set, we notice that the outer disk of planetesimals might have been too close to the outermost ice giant planet at the beginning of the simulation. Indeed, the outer ring of planetesimals is located between 21 and 30 AU and the outermost planet starts at $\simeq$ 22 AU. This has the effect of completely destabilizing the outer disk in less than 1 Myr. Consequently, outer planetesimals are already scattered towards the inner and outer solar system before the onset of the giant planet instability. In the context of our study nevertheless, this is of no great significance since our main goal is to show a \textit{range} of possible timings for the cometary bombardment relative to the accretion of the Earth.

Out of the 13 successful simulations, 7 started from 20 $\textit{big}$ embryos ($\textit{big}$ set) and 6 started from 40 $\textit{small}$ embryos ($\textit{small}$ set). Hence, it cannot be decided either which initial conditions were the best between the $\textit{big}$ and $\textit{small}$ sets. However, the accretional growth timing of Earth best fits geophysical constraints \cite{Kleine,Rudge} in the $\textit{small}$ initial configuration (with 30 Myr $< t_{growth} <$ 90 Myr), with respect to the $\textit{big}$ scenario ($t_{growth} <$ 10 Myr in most cases). Two simulations from the $\textit{big}$ set (CL2/big/2 and NE6/big/1) have unrealistically short growth timings, with more than 95\% of their respective masses accreted in less than 1 Myr (see Table \ref{table2}). However, as the timing of Earth’s growth does not influence the cometary bombardment, it is still interesting to study the pattern of their corresponding cometary influx.

\begin{figure}[h!]
  \centering
  \includegraphics[width=11.5cm]{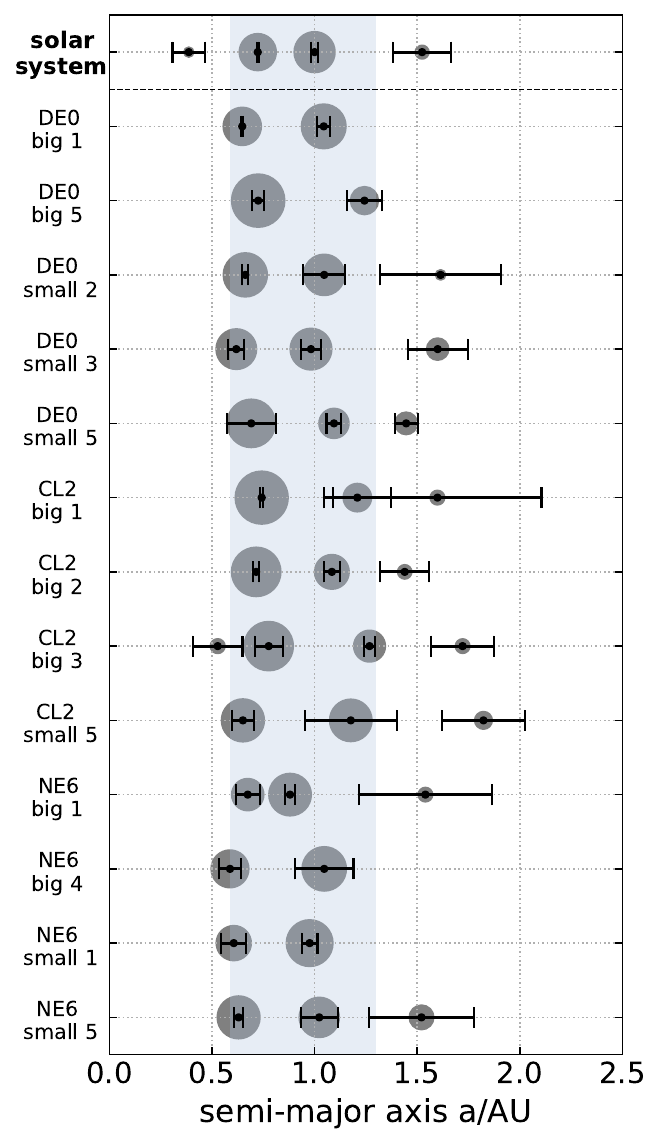}
  \caption{\small Comparison of the terrestrial planets analogs in their final states (at t = 100 Myr) from all successful simulations (according to the list of criteria presented in Section \ref{success}), with the current state of the inner Solar System. The symbol sizes are proportional to the mass of the bodies and the error bars indicate the variation in heliocentric distance given by the perihelion and aphelion of the orbit. The area marked in blue defines the semi-major axes range for which there must be two planets, the Venus and Earth analogs respectively. Before this range are the Mercury analogs and after are the Mars analogs.}
  \label{inner}
\end{figure}

In Figure \ref{inner}, we display the final outcome of each of the 13 successful simulations, i.e. the state (number, relative mass, semi-major axis and eccentricity) of the inner planets at t = 100 Myr. 

\begin{figure}[h!]
  \centering
  \includegraphics[width=11cm]{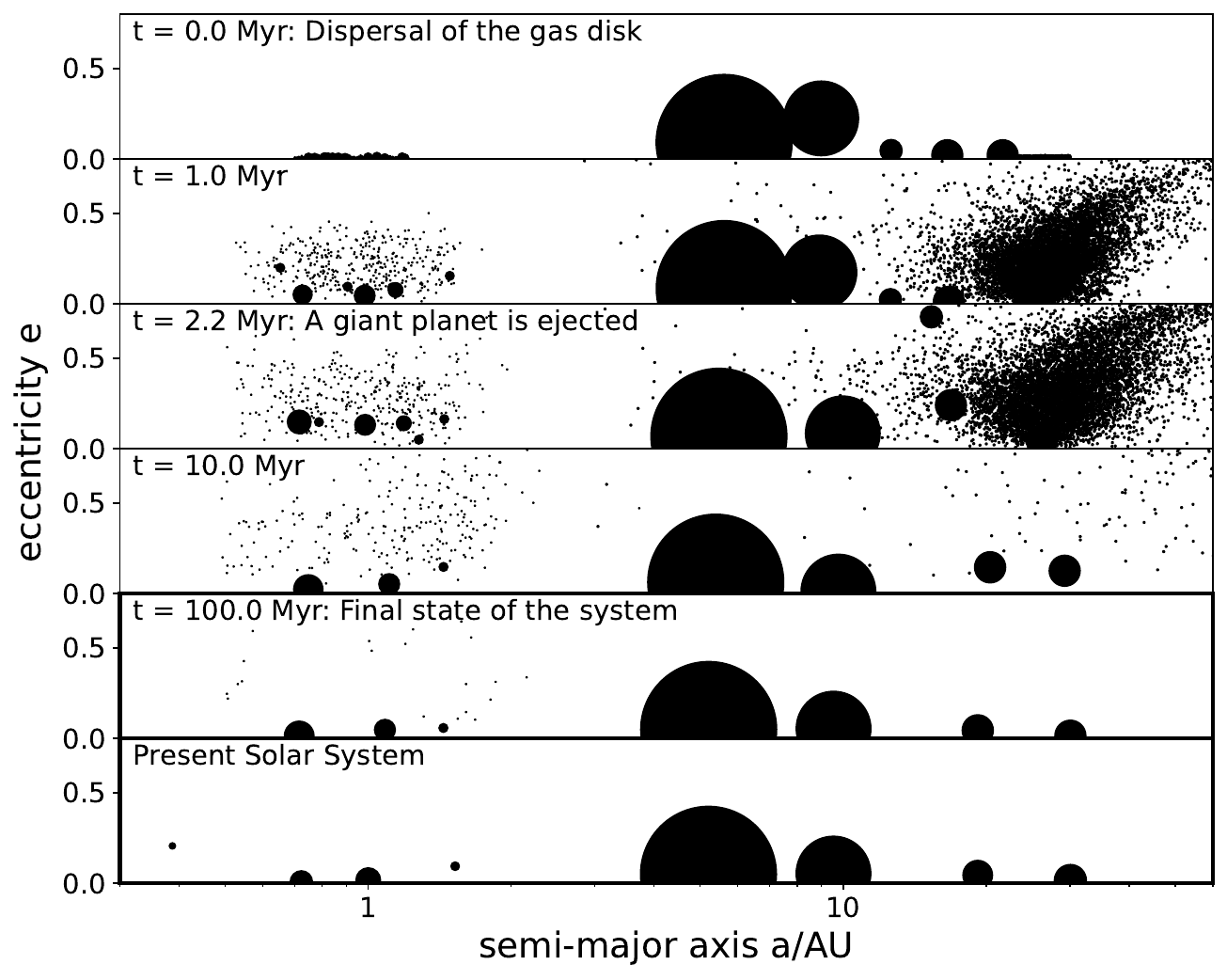}
  \caption{\small Simulation CL2/big/2. The symbol sizes are proportional to the mass of the bodies; for the giant planets a fixed size was chosen so as not to fill most of the panels.}
  \label{CL2}
\end{figure}

9 of the 13 successful simulations include a Mars analog ($1.3 < a < 2$ AU and mass $\lesssim$ $\frac{M_{Earth}}{3}$), and only one of them includes a Mercury analog (a $<0.6$ AU and mass $\lesssim$ $\frac{M_{Earth}}{4}$). The Mercury analog has a mass of 0.123 $M_{Earth}$, which is about 2 times larger than the actual Mercury. The Mars analogs have a median mass of 0.121 $M_{Earth}$, which is close to the actual Mars mass (0.107 $M_{Earth}$), with a small variability among the results. In 8 of these 13 good simulations, the Venus analog is larger than the Earth analog. The median mass of the Venus and Earth analogs is 1.101 $M_{Earth}$ and 1.015 $M_{Earth}$ respectively. If we take a closer look at the differences of outcomes in \textit{DE0}, \textit{CL2} and \textit{NE6} sets, we find that the average mass ratios between Venus and Earth analogs are 1.82, 2.26, and 0.75, respectively. For \textit{NE6}, Venus is smaller than the Earth by a ratio that is close to the real one ($\simeq$ 0.815). In all other cases, Venus tends to be more massive than the Earth. The early timing of the instability could play a role in this, in the way Mars' and Earth's feeding zone are depleted. One might also notice that \textit{DE0} and \textit{NE6} do a better job of getting a tight spacing between the Venus and Earth analogs.

The evolution of the terrestrial planets analogs are stable after the first 100 millions years, except in simulation CL2/big/1. Indeed, the Mars analog in this simulation is very eccentric and its orbit crosses the Earth analog's orbit, which could lead to an ejection or a collision.

Figure \ref{CL2} shows the temporal evolution of simulation CL2/big/2, and a comparison with the present solar system. Before the onset of the giant planet instability and the ejection of the fifth planet, the Neptune analog already migrates outwards and the outer ring of planetesimals is destabilized. Most of these planetesimals are ejected outwards but some of them are ejected inwards and reach the inner solar system. By the end of the instability, the outer ring of planetesimals is strongly depleted. After 100 million years of simulation, the inner solar system ends up with 3 stable rocky planets following similar orbits to that of the actual Venus, Earth and Mars. In this particular simulation, the Venus and Earth analogs grew too rapidly, but this is related to the initial size of the embryos in the $\textit{big}$ scenario. 

\begin{figure}[h!]
  \centering
  \includegraphics[width=11cm]{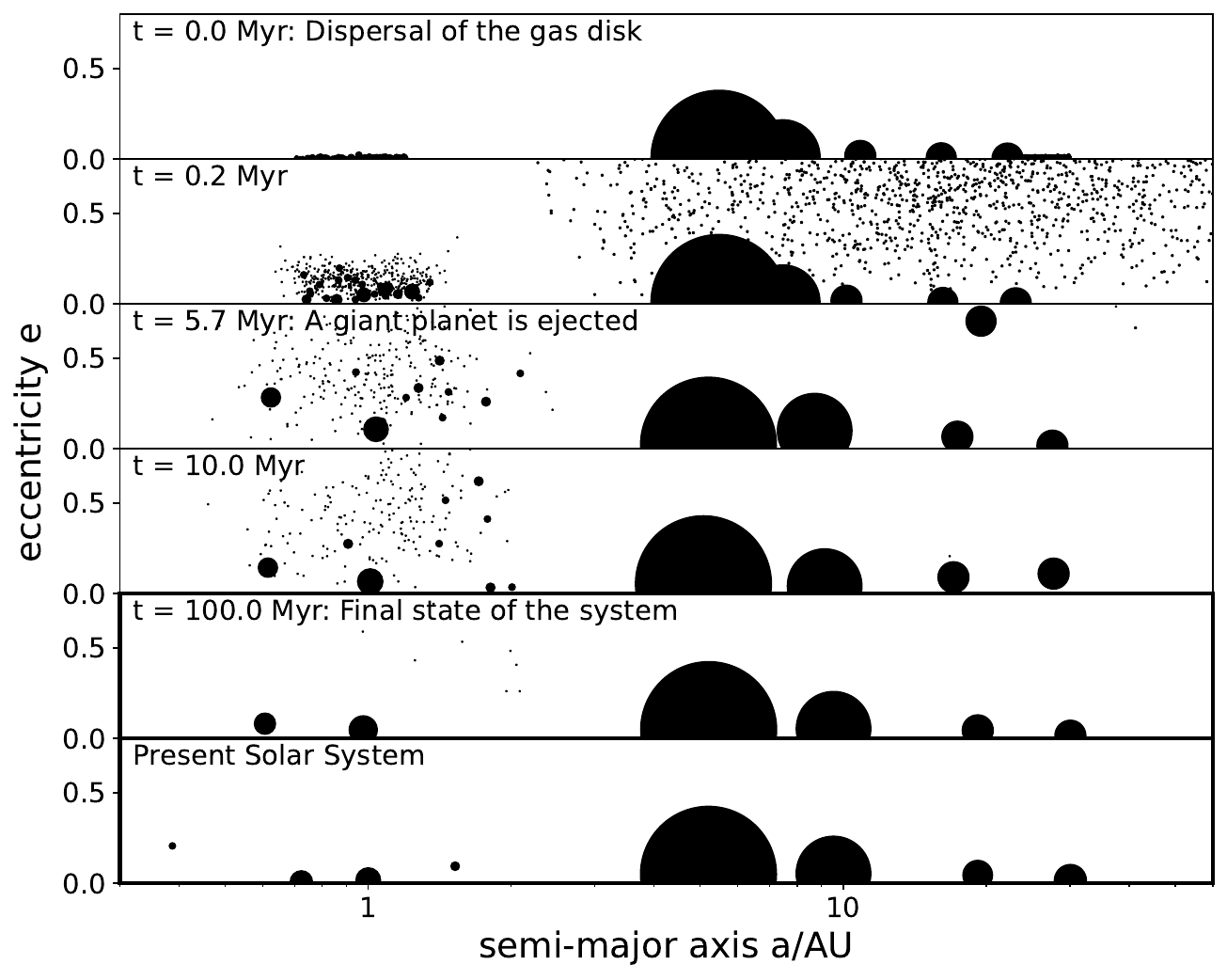}
  \caption{\small Simulation NE6/small/1. The symbol sizes are proportional to the mass of the bodies; for the giant planets a fixed size was chosen so as not to fill most of the panels.}
  \label{NE6}
\end{figure}

In Figure \ref{NE6}, we show the outcome of a simulation starting from smaller embryos (NE6/small/1). The inner solar system ends up with 2 stable rocky planets but with a more realistic growth timing for the Earth. Indeed, the last giant impact happens at $t = 38.1$ Myr and the impactor has a mass of 0.12 $M_{Earth}$. The Earth analog accretes 0.02\% of its final mass after this giant collision. This is close to what the dynamical and geochemical constraints require on Earth with the last giant impact taking place between 30 and 100 Myr after CAIs \cite{Kleine}, an impactor of mass $\simeq$ 0.1-011 $M_{Earth}$ \cite{Canup2001}, and 0.05\% of Earth final mass accreted after that episode \cite{Jacobson}. Regarding the instability, we see that the outer disk of planetesimals has already been destabilized at the very beginning of the simulation. As explained above, this is probably because the Neptune analog was located too close to the disk in the initial conditions of the \textit{NE6} set.

\begin{figure}[h!]
  \centering
  \includegraphics[width=11cm]{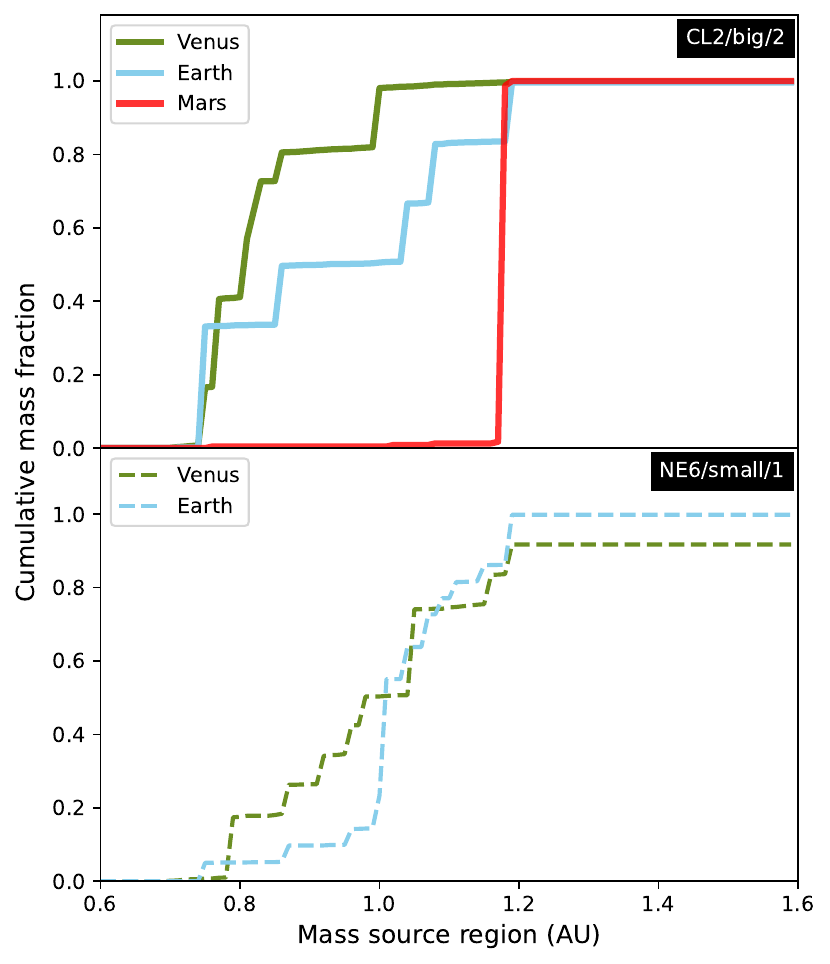}
  \caption{\small Feeding zones of the terrestrial planets in simulations CL2/big/2 (top panel) and NE6/small/1 (bottom panel). They are represented by their cumulative mass as a function of the initial semi-major axis of their component embryos.}
  \label{feeding}
\end{figure}

We calculate the feeding zones of the terrestrial planets in our simulations and see how they vary through time (cf. Figure \ref{feeding}). Feeding zones have significant consequences for the volatile inventory of planets. However, they are difficult to predict given that the last stages of planet formation are stochastic \cite{Kaib2015}. We find that Earth and Venus analogs have similar feeding zones, whereas Mars mostly accretes material from the outer edge of the terrestrial planetesimal ring. This result is consistent with \citet{Izidoro2022}'s conclusions in the framework of the \textit{annulus} model.

\subsection{Collision rate calculations}
\label{results_coll}
For each simulation and following the method described in section \ref{coll_cal}, we compute the intrinsic collision probability between each pair of comet and Earth embryo (main embryo or future Earth constituent embryo) at each timestep. This probability can be converted in a number of collisions occurring during a certain time interval. For each timestep, we add up the number of collisions with a comet of all the main and future Earth constituent embryos. Indeed, the Earth constituent embryos that haven't merged with the main Earth analog embryo yet, may also collide with comets and bring cometary material to the fully-formed Earth. 

\begin{figure*}[h!]
  \includegraphics[width=\linewidth]{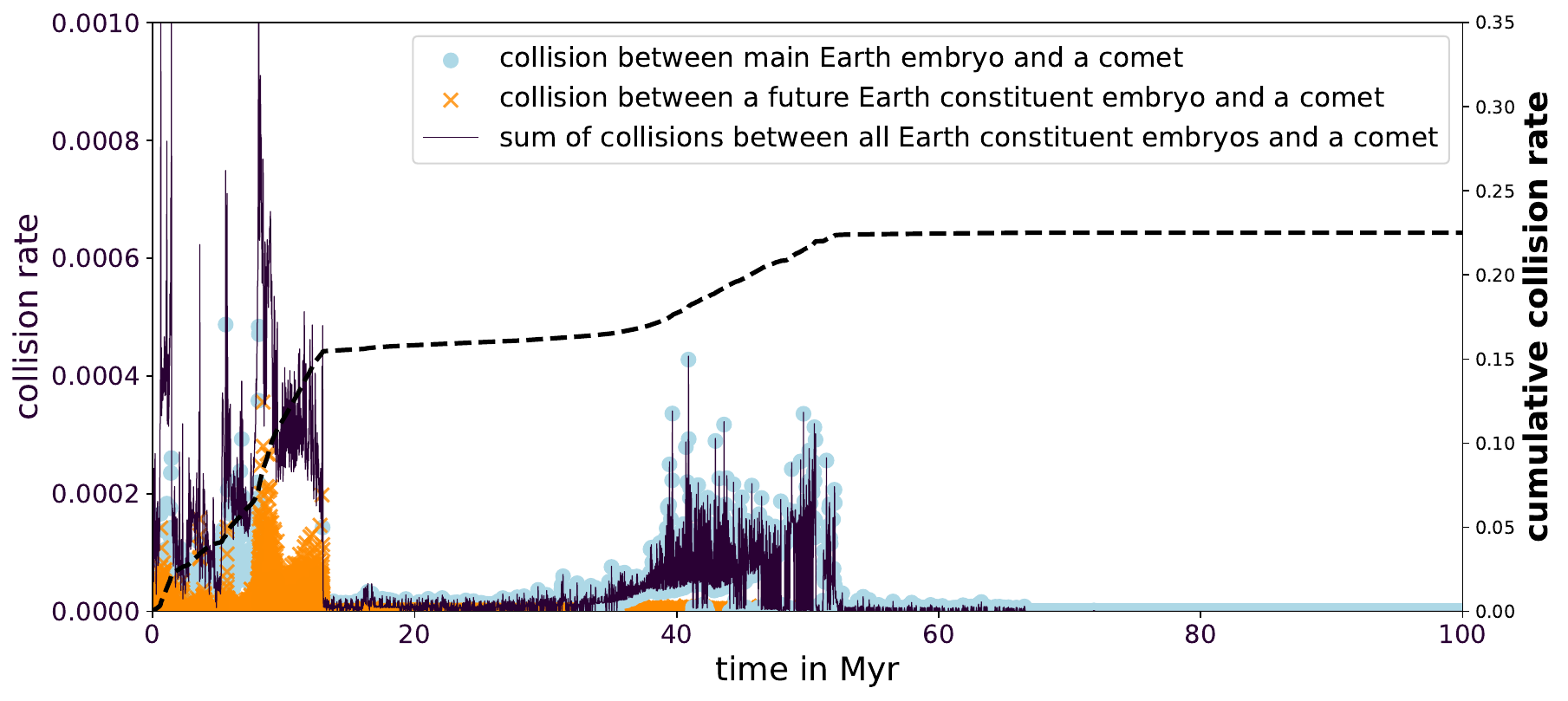}
  \caption{\small Temporal evolution of the collision rate between comets and Earth constituent embryos in the simulation CL2/big/2. The giant planet instability happened at t $\simeq$ 2 Myr. For simplicity, we use the term 'Earth constituent embryos' even when we refer to Earth constituent planetesimals. The presence of orange crosses up to a certain point shows that the main Earth analog embryo is still accreting its Earth constituent embryos. These future Earth constituent embryos are also prone to colliding with a comet. Hence, the cometary mass they accrete before merging with the main Earth embryo will be integrated in the fully-formed Earth. The sum of the cometary contributions of all these Earth embryos is shown in dark purple. The cumulative collision rate is represented with the dashed line on the right axis. In 100 millions years, the probability of collision between an Earth constituent embryo and a comet is about 23\%.}
  \label{nb_collision}
\end{figure*}

\begin{figure}[h!]
  \centering
  \includegraphics[width=12cm]{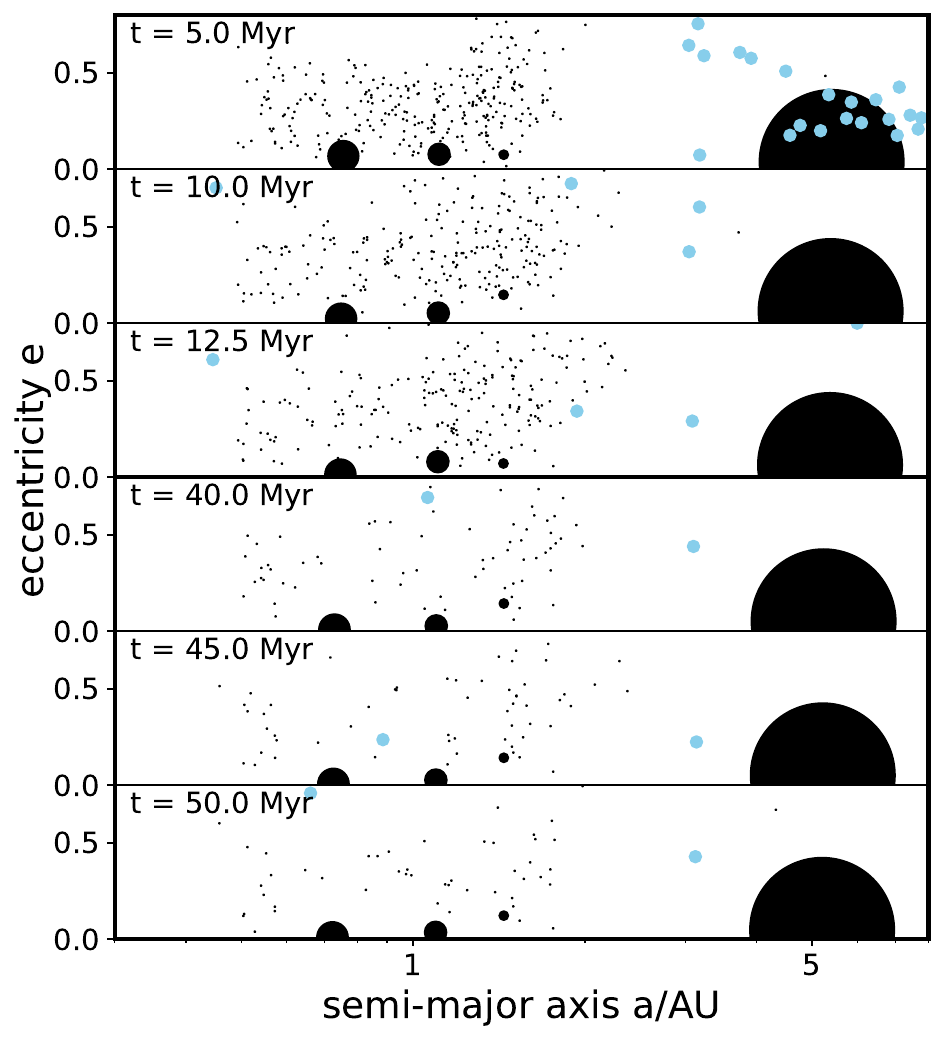}
  \caption{\small Simulation CL2/big/2 with a zoom on the inner solar system. Different timings at which the peaks in collision rate are particularly high in this simulation are displayed. The symbol sizes are proportional to the mass of the bodies, except for the comets (in blue) for which the size has been exaggerated and Jupiter (around 5 AU) for which a fixed size was chosen so as not to fill most of the panels.}
  \label{inner_comet}
\end{figure}

Figure \ref{nb_collision} shows the temporal evolution of this rate of collisions with comets for simulation CL2/big/2. This particular simulation reveals that, even though the giant planet instability happens at t $\simeq$ 2 Myr, the number of collisions with comets shows important peaks up to t $\simeq$ 13 Myr and between $\simeq$ 40 and 50 Myr. A closer look into the dynamical evolution of this simulation reveals that the peaks in collision rate before t $\simeq$ 13 Myr and between $\simeq$ 40 and 50 Myr are both due to only one or two comets orbiting close to the terrestrial embryos (Figure \ref{inner_comet}). This shows that cometary accretion on terrestrial embryos could have included a stochastic component.  More importantly, it can be concluded that the contribution of cometary material on Earth might have been delayed with respect to the timing of the instability. 

The cumulative collision rate, which is smaller than one ($\simeq$ 0.23 in simulation CL2/big/2 for example, cf. Figure \ref{nb_collision}), justifies the need for a probabilistic model for the collision rate calculations, rather than directly counting the number of collisions with comets from the N-body simulations. The collision rate we obtain for each simulation depends to a large extent on the number of comets considered for the initial conditions. If we increased the number of outer solar system objects, the intrinsic probability of collision between an Earth embryo and a comet, and in turn the collision rate, would also increase. However, this increment would be limited by the radius of each comet being smaller, given a total mass of the outer disk set to 25 $M_{Earth}$, which would also reduce the volume where the collision can take place. In our simulations, the 10000 comets are unrealistically large in order to balance for their \textit{small} population. Hence, estimating the cometary mass hitting Earth (see Section \ref{cmc}) or the collision rate normalized to the number of comets (see Section \ref{ip}), instead of the collision rate for all 10000 comets, is more relevant in the scope of our study.

While the assumption of each comet being $\simeq 1.5 \times 10^{25}$ g (about the mass of Pluto) appears to be highly unrealistic, it has been shown that the combined mass of Pluto-mass objects in the original outer disk could represent 10\% to 40\% of the disk mass \cite{Nesvorny2016}.  It is estimated that the outer disk below 30 AU would have contained 1000–4000 Plutos initially \cite{Nesvorny2016}.  More recently, \citet{Kaib2021} argued that the number of Pluto-mass bodies in the primordial disk was smaller than $\simeq$ 1000 . Even though the precise number of those massive bodies could still be debated, it means that a non-negligible part of the comets in our simulations is representative of the original outer disk. The remaining part should naturally contain a large amount of much smaller bodies.  In any case,  the Pluto-mass objects carried a significant part of the mass in the early outer disk and could be responsible for the stochastic component of the cometary bombardment in the inner solar system. The contribution of the smaller comets would add a smoother component to the bombardment. The latter has been studied by \citet{Nesvorny2023}. They have modeled the cometary impacts on Earth through time (after the Moon-forming impact) using the distribution of comets from \cite{Nesvorny2017b}. The size distribution of the comets considered in their study is mainly comprised between 10 and 100 km, with a very steep slope for comets with diameters $>$ 100 km (i.e. for comets with mass $ \gtrsim 2.6 \times 10^{20}$ g). So they consider much smaller bodies than we do in this study. They calculate the collision probability between these \textit{small} comets and a target body at 1 AU (the Earth) and obtain a smooth profile for the cometary bombardment after the Moon-forming impact. As their approach is complementary to ours, we  compare our quantitative results in Section \ref{discu_mass}.

\subsubsection{Cometary mass calculations}
\label{cmc}
\begin{figure*}[h!]
  \includegraphics[width=\linewidth]{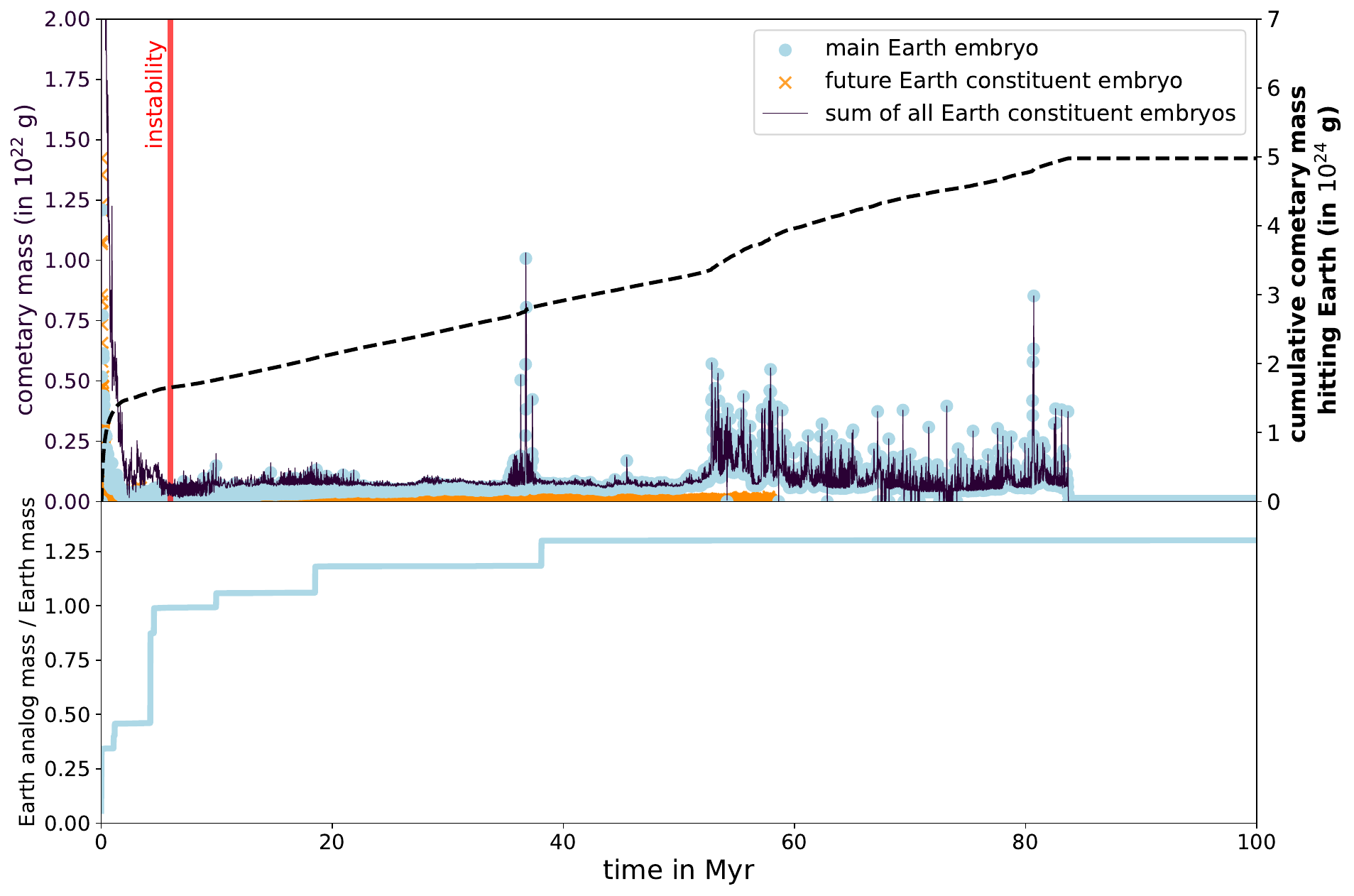}
  \caption{\small Temporal evolution of the cometary mass hitting Earth constituent embryos in the simulation NE6/small/1. The upper part of the figure shows the cometary mass with the dark purple line on the left axis and the cumulative cometary mass with the dashed black line on the right axis. The lower part of the figure shows the accreting mass of the main Earth embryo with a last giant impact at t $\simeq$ 40 Myr. For simplicity, we use the term 'Earth constituent embryos' even when we refer to Earth constituent planetesimals. The presence of orange crosses up to t $\simeq$ 60 Myr on the upper part of the plot implies that small planetesimals continued to collide with the Earth in a late accretion phase, to eventually reach a final mass of 1.3 $M_{Earth}$.}
  \label{commass}
\end{figure*}

As detailed in Section \ref{coll_cal}, we infer the cometary mass hitting an Earth analog constituent embryo by multiplying its number of collisions with a comet by the mass of one comet in our simulations ($\simeq 1.5 \times 10^{25}$ g; cf. section \ref{IC}). This cometary mass hitting Earth constituent embryos as a function of time in our simulation NE6/small/1 is represented in Figure \ref{commass}. The running mass of the Earth is also shown, with a last giant impact happening at t = 40 Myr. Interestingly, this result demonstrates that it is possible that most cometary mass was brought to Earth \textit{after} it had finished forming.

The high peak in cometary mass at the very beginning of Figure \ref{commass} is due to the outer disk of planetesimals being destabilized in the first Myr, even before the onset of the giant planets instability. The reasons for this have been explained in section \ref{outcomes} and is most probably due to the initial conditions of the simulation. This appears in every instability case but the \textit{NE6} set is where it is the most obvious. In the following, we \textit{normalize} the cometary mass brought to Earth to the timing of the instability. In other words, we only take into account the contribution of comets in Earth constituent embryos starting from the onset of the giant planet instability. This has implications on the values obtained for the cometary mass brought to Earth, which could be slightly underestimated, especially in the instability case \textit{NE6} (see Table \ref{table2}). 

\begin{figure*}[h!]
  \includegraphics[width=\linewidth]{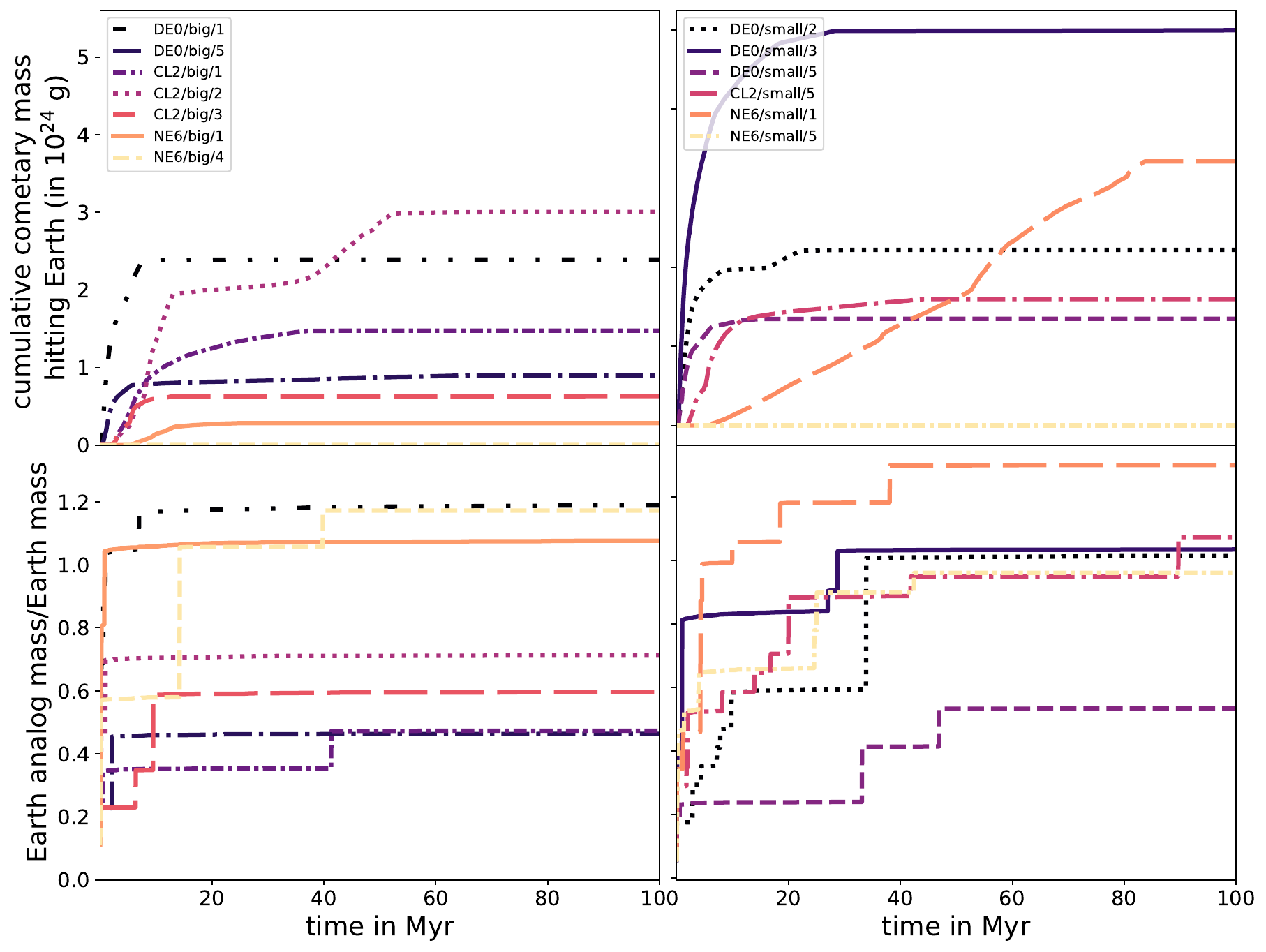}
  \caption{\small Comparison of the cometary bombardment timings (top panels) relative to Earth analog growth (bottom panels) for each of the successful simulations. The left panels consider the simulations which started with Mars-sized embryos (the \textit{big} set) in the inner solar system; and the right panels show simulations which started with a few lunar masses (the \textit{small} set).}
  \label{comp2}
\end{figure*}

Figure \ref{comp2} shows a comparison of all the cumulative cometary masses hitting Earth as a function of time for all the successful simulations (top panels). These functions are \textit{normalized} to the timing of the instability, which explains why they start increasing almost directly in simulations \textit{DE0}, but only after t = 2 and 6 Myr in simulations \textit{CL2} and \textit{NE6} respectively. The bottom panels display the accreting mass rates of the corresponding Earth analogs. The moment of the last giant impact (or Moon-forming impact) happens between $\simeq$ 0 and 45 Myr in the \textit{big} set (left panel) and between $\simeq$ 30 and 90 Myr in the \textit{small} set (right panel), which is a little bit more realistic in terms of geophysical constraints \cite{Kleine}. It should be noted that simulations CL2/big/2 and NE6/big/1 have very unrealistic growth timings, but it does not have any influence on the absolute timing of the cometary bombardment. The cometary bombardment duration is highly variable and can last for less than 10 Myr as well as $\simeq$ 80 Myr. The exact values appear in Table \ref{table2}.

\begin{table*}[t!]
\begin{tabularx}{\linewidth}{lYYYYYYYY}
\hline
      & Normalized \newline cometary bomb. \newline mass (g) & Fraction of com. mass wrt. Earth & Cometary bomb. mass after last giant impact (g) & Cometary bomb. duration (Myr) & Last giant impact time (Myr) & Last giant impact mass ($M_{Earth}$) & Late accretion mass ($M_{Earth}$) \\
 \textbf{constraints} & \textbf{/} & \textbf{$<$0.5\%} & \textbf{/} & \textbf{/} & \textbf{[30-100]} & \textbf{$\simeq$ 0.1} & \textbf{$\simeq$0.005} \\ \hline
 \textit{DE0}/big/1 & 2.39e+24 & 0.034\% & 9.55e+22 & 15.5 & 6.9 & 0.11 & 0.026 \\ 
 \textit{DE0}/big/5 & 9.00e+23 & 0.033\% & 3.98e+23 & 7.5 & 2.1 & 0.23 & 0.009 \\ 
 \textit{DE0}/small/2 & 2.22e+24 & 0.037\% & 3.71e+19 & 7.1 & 33.9 & 0.41 & 0.007 \\
 \textit{DE0}/small/3 & 5.00e+24 & 0.081\% & 8.04e+21 & 7.5 & 28.8 & 0.12 & 0.006 \\
 \textit{DE0}/small/5 & 1.35e+24 & 0.042\% & 6.36e+18 & 50.7 & 46.9 & 0.12 & 0.001 \\ \hline
 \textit{CL2}/big/1 & 1.48e+24 & 0.052\% & 2.20e+21 & 1.1 & 41.3 & 0.12 & 0.001 \\
 \textit{CL2}/big/2 & 3.00e+24 & 0.071\% & 3.21e+24 & 29.3 & 0.9 & 0.24 & 0.015 \\ 
 \textit{CL2}/big/3 & 6.32e+23 & 0.018\% & 3.69e+22 & 9.8 & 9.4 & 0.24 & 0.009 \\ 
 \textit{CL2}/small/5 & 1.60e+24 & 0.025\% & 3.12e+15 & 11.7 & 89.7 & 0.12 & 0.0 \\ 
 \hline
 \textit{NE6}/big/1 & 2.85e+23 & 0.004\% & 1.35e+24 & 13.7 & 0.7 & 0.23 & 0.034 \\ 
 \textit{NE6}/big/4 & 4.42e+21 & $<$0.001\% & 0.0 & $\simeq$0.0 & 39.8 & 0.12 & 0.0\\
 \textit{NE6}/small/1 & 3.34e+24 & 0.043\% & 2.13e+24 & 78.2 & 38.1 & 0.12 & 0.002 \\
 \textit{NE6}/small/5 & 0.0 & 0.0\% & 0.0 & 0.0 & 42.4 & 0.06 & 0.001 \\
 \hline
\end{tabularx}
\caption{\small Summary of all successful simulations. The values specific to the solar system are displayed for comparison. Second column is calculated starting from the moment of cometary bombardment. Third column shows the fraction of cometary mass with respect to the Earth analog mass in the simulation.}
\label{table2}
\end{table*}

\subsubsection{Collision rate of each comet with Earth}
\label{ip}
In this section , we take simulation CL2/big/2 as a case study to estimate collision rates of each comet with the Earth. In this particular scenario, the cumulative collision rate between Earth's constituent embryos and 10000 comets is $\simeq$ 0.23, over the 100 millions years of our simulation. This means that the collision rate of each comet with Earth, over the same period of time, is equal to $\simeq$ 2.3 $\times 10^{-5}$. 

Considering the fact that comets in our simulations are Pluto-mass objects and there might have been between 1000 and 4000 of these objects in the primordial Kuiper Belt \cite{Nesvorny2016}, the collision rate between a Pluto object and an Earth embryo in the early solar system would range between $\simeq$ 0.02 and 0.09. 

The Kuiper Belt objects populate as $N(>d)$ $\simeq$ $d^{-q}$ , with $4.5 < q < 7.5$, for objects with a diameter d$>$100 km \cite{Fraser2014, Morby2021}. Hence, the number of objects larger than half the size of Pluto, could be 20 to 200 times more numerous than Pluto objects. The collision rate between these objects and Earth would thus increase accordingly, reaching values ranging between 0.4 and 18 collisions. These smaller - but still very massive - objects could have played an important role in the stochastic component of the bombardment.

\section{Discussion}
\subsection{Timing of cometary accretion with respect to Earth's growth}
We can distinguish three possible scenarios for the relative timing of cometary bombardment and Earth's accretion. In the first one, there is (almost) no cometary bombardment after the last giant impact. This is the most prevalent scenario among all our successful outcomes (10 out of 13). Simulations CL2/big/1 or DE0/small/2, for example, follow this particular evolution. For those Earth analogs, a \textit{cometary} isotopic signature for the atmospheric noble gases could hardly be explained. In addition, we should find a signature of comets in the Earth's mantle. An hypothesis that should be investigated in future works is that the cometary volatile material could simply not be trapped efficiently in the mantle. 

In the second scenario, a significant part of the cometary bombardment happens after the last giant impact. This is particularly the case for simulations DE0/big/5 and NE6/small/1. For those Earth analogs, there should be a cometary signature of noble gases in the atmosphere \textit{and} in the mantle.  

In the third and last scenario, there is almost no cometary bombardment on Earth until after the Moon-forming impact. So, in principle, we should find a cometary signature in the atmosphere of those Earth analogs, but not in their mantle. Simulations CL2/big/2 and NE6/big/1 exemplify this potential scenario. 

If we consider noble gas constraints on Earth, it appears that the mantle xenon is chondritic \cite{Peron, Broadley}, while atmospheric xenon can be separated into a chondritic compoment ($\simeq$ 78\%) and a cometary one ($\simeq$ 22\%) \cite{Marty}. It is not possible to rule out a small cometary contribution to the mantle of the Earth (see Section \ref{gc}), so the most plausible scenario for the actual Earth is either the second or the third one. It is important to note that the simulation examples we have for the third scenario (CL2/big/2 and NE6/big/1) always involve an extremely rapid, and thus unrealistic, formation of the Earth analogs ($t_{growth}$ $<$ 1 Myr). This is most probably the major reason why comets are exclusively delivered after the last giant impact in these simulations. With realistic growth timings, simulation CL2/big/2 would probably fall into the second scenario and simulation NE6/big/1 would fall into the first scenario. A bigger sample in simulation outcomes is necessary to validate this possible dynamical evolution. However, in a first approach, it appears unavoidable to bring cometary material to the mantle of the Earth in the context of an \textit{Early Instability} scenario. 

The second scenario might also match Earth's accretion history, with a cometary delivery \textit{before} (as well as after) the last giant impact and Earth's core closure. Simulation DE0/big/5 involves a very rapid formation of the Earth analog ($t_{growth}$ $\simeq$ 2 Myr) and would fall into the first scenario with a realistic growth timescale. In contrast, simulation NE6/small/1 satisfy many dynamical and geochemical constraints with a last giant \textit{Mars-sized} impact $\simeq$ 40 Myr after $t_{0}$ and 0.2\% of mass accreted after this impact. Regarding the cometary bombardment in this simulation, $\simeq$ 1/3 of the cometary material is delivered \textit{before} the last giant impact, and $\simeq$ 2/3 \textit{after}. This could be consistent with a 10\% cometary contribution in the mantle allowed by the fact that chondritic Xe and U-Xe (mixture of chondritic and cometary Xe) can hardly be distinguished as a progenitor for the mantle due to measurement uncertainties (see Section \ref{gc}). As also detailed in Section \ref{gc}, comets would have contributed to the deep mantle's Kr budget during the earliest phase of Earth's growth \cite{Peron2021}. In that case, the cumulative cometary mass hitting Earth as a function of time (as shown in Figure \ref{comp2}) would be an upward slope in the first millions years, it would then reach a plateau until $\simeq$ 40 Myr for example, and there would be another upward slope followed by a plateau; just like in simulation CL2/big/2. If future geochemical studies also find a deficit in $^{86}$Kr in the upper mantle, suggesting that a few comets contributed to the composition of Earth's upper mantle too, the cumulative cometary mass hitting Earth as a function of time could be similar to that of simulation NE6/small/1.

Overall, and given the issue with the very early formation of Earth analogs in the third scenario, the second scenario appears the most likely. These results support the claim that comets have contributed to the mantle early in the Earth's accretion history, but that cometary material could also have been delivered after the Earth had finished forming.

\subsection{Total cometary mass accreted by the Earth}
\label{discu_mass}
We assume that there is no loss of volatile elements to space during collisions between comets and terrestrial embryos. This assumption is discussed in Section \ref{limitations}. The values we obtain for the cumulative cometary mass accreted by our Earth analogs through time (during the first 100 Myr) range between 4 $\times$ $10^{21}$ and 5 $\times$ $10^{24}$ g, which also amounts to between 6.7 $\times$ $10^{-7}$ $M_{Earth}$ and 8.4 $\times$ $10^{-4}$ $M_{Earth}$ (see Table \ref{table2}). 

A rough estimation of the expected cometary contribution on Earth can be obtained by comparing it to the carbonaceous chondrites contribution on Earth. \citet{Burkhardt2021} show that CC material could have contributed from $\simeq$ 0 to 10 \% of Earth by mass, with a peak at $\simeq$ 4\%. More recently, a $\simeq$ 6\% contribution has been proposed by \citet{Savage2022}, using Zn isotope anomalies. Based on the D/H ratio in the present-day oceanic water, it has been shown that the contribution of comets cannot exceed 10 \% of the contribution of carbonaceous chondrites \cite{Morbidelli2000}. Hence, comets could not represent more than about 0.5\% of Earth mass, which amounts to 5 $\times$ $10^{-3}$ $M_{Earth}$ $\simeq$ 3 $\times$ $10^{25}$ g. 

These values for the cometary mass on Earth are based on the total contribution since $t_{0}$ (or almost $t_{0}$ with the normalization to the timing of the instability). In what follows, we estimate the expected mass brought by comets \textit{after} the Moon-forming impact in order to compare it with geochemical constraints in the atmosphere of the Earth. We find that the cometary mass accreted by our Earth analogs after the last giant impact ranges between 3 $\times$ $10^{15}$ and 3 $\times$ $10^{24}$ g (except in two cases where it is zero), or between 5.0 $\times$ $10^{-13}$ $M_{Earth}$ and 5.0 $\times$ $10^{-4}$ $M_{Earth}$. The lower limit of this range is reached in simulation \textit{CL2/small/5} where the last giant impact happens at 89.7 Myr after $t_{0}$, whereas the upper limit is reached in simulation \textit{CL2/big/2} where the timing of the last giant impact is unrealistically early. However, in simulation \textit{NE6/small/1}, the last giant impact happens at 38.1 Myr after $t_{0}$, which is very realistic, and 2$\times$ $10^{24}$ g of cometary material is brought to the Earth analog after this event. 

\citet{Marty2016} argue that a cometary contribution between 3 $\times$ $10^{21}$ and 6.5 $\times$ $10^{23}$ g (or between 5 $\times$ $10^{-7}$ and 1 $\times$ $10^{-4}$ $M_{Earth}$) would have been necessary to supply all atmospheric $^{36}$Ar.

Alternatively, and based on the amount of Kr residing in the Earth surface reservoir, \citet{Bekaert2020} estimate the maximum amount of water that would have been brought by comets along with cometary Kr after the Moon-forming impact to $\simeq$ 1.95 $\times$ $10^{21}$ g. From this value, they derive the total mass of comets accreted by the Earth after the last giant impact and obtain a value of $\simeq$ 9.76 $\times$ $10^{21}$ g (or $\simeq$ 1.6 $\times$ $10^{-6}$ $M_{Earth}$). 

Either way, an impact of one single comet with a 100-km radius and density of 0.5 g/cm$^{3}$, assuming that it is perfectly spherical and that volatile elements are conserved during the collision, could supply $\simeq$ 2 $\times$ $10^{21}$ g of material to Earth.  An impact of a comet with a 1000-km radius, and same density, could supply $\simeq$ 2 $\times$ $10^{24}$ g of material to Earth. Therefore, only a small number of collisions may have been necessary to account for the cometary contribution to Earth. Our simulations show that it is possible that enough cometary mass has been brought to Earth after it had finished forming, through big impactors, in order to explain the noble gas content in the Earth's atmosphere.  

Because the primordial outer disk of planetesimals contained a non-negligible portion of Pluto-mass objects \cite{Nesvorny2016, Kaib2021}, the cometary bombardment on Earth probably included both a stochastic component - dictated by Pluto-mass objects -, and a smoother component - dictated by smaller comets. As explained in details in Section \ref{results_coll}, \citet{Nesvorny2023} have calculated the cometary mass reaching Earth after the Moon-forming impact, considering a smaller size distribution of comets (instead of Pluto-mass objects like we do) and neglecting the stochastic component.  They find that, assuming a Moon-forming impact at t $\simeq$ 50 Myr after the birth of the solar system, the Earth would have accreted $\simeq 1.5\times 10^{22}$ g of cometary material \textit{after} the Moon-forming impact. Their results also suggest that the time gap between the instability and the Moon forming event should be shorter than 50 Myr, in order to explain the noble gases content in the Earth's atmosphere. The same result holds for our simulations under certain assumptions. Indeed, it is also possible that the noble gases delivered by the comets were just not retained efficiently in the mantle, and were released to the atmosphere even before the last giant impact. This hypothesis needs to be tested before constraining the timing of the Moon-forming event relative to the timing of the instability. 

\subsection{Comparison with Venus and Mars analogs}
We calculate the intrinsic collision probability between each pair of comet and Venus analog embryo, and each pair of comet and Mars analog embryo, at each timestep of our 100 Myr simulations. Subsequently, we obtain the temporal evolution of the cometary mass that has been supplied to these terrestrial planets analogs and compare it with our Earth analogs.

\begin{figure}[h!]
  \centering
  \includegraphics[width=9cm]{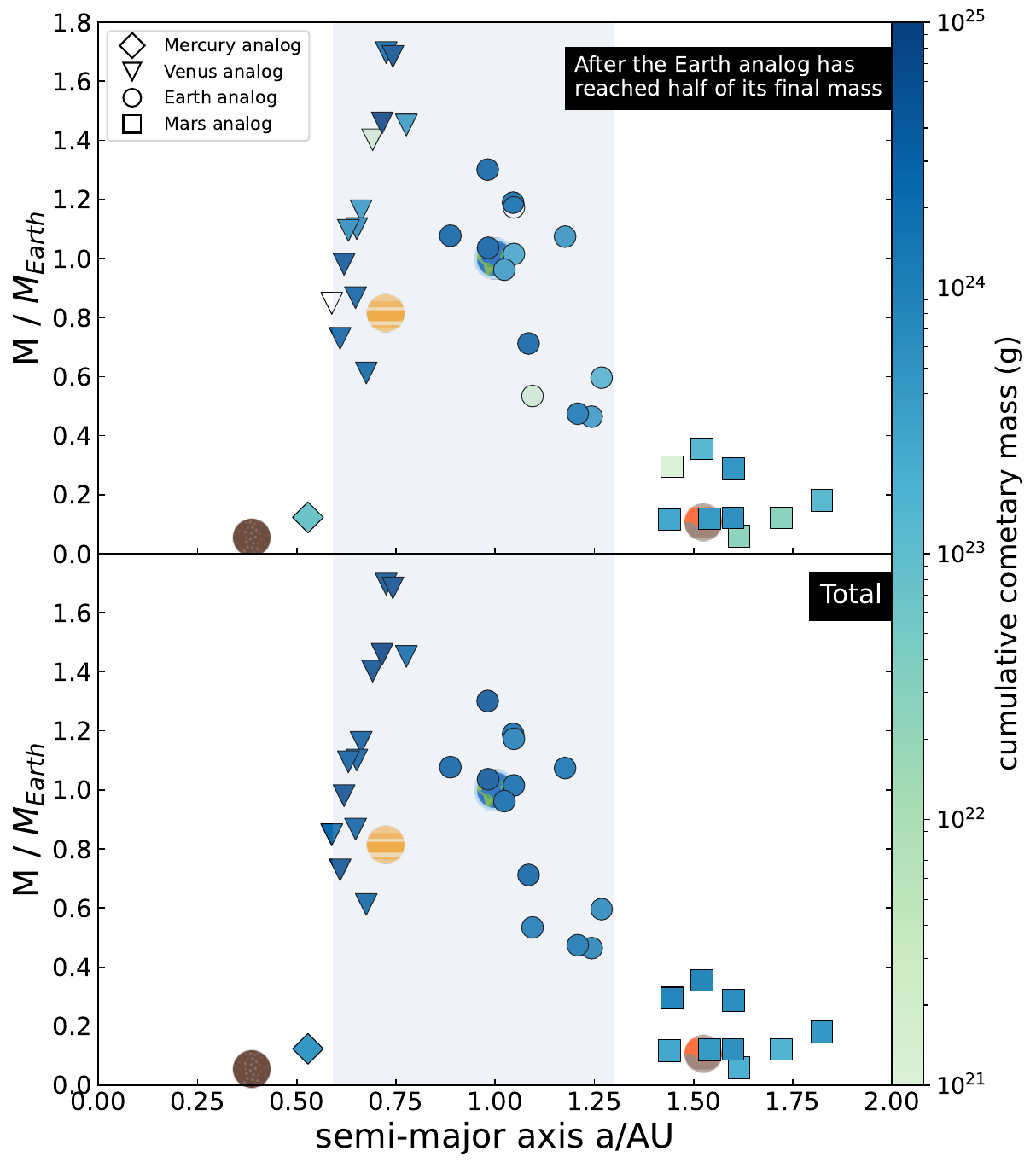}
  \caption{\small Semimajor axis–eccentricty distribution of the terrestrial planets analogs that formed in all our successful simulations, and the corresponding cumulative cometary mass that has been delivered to them after the Earth analog has reached half of its mass (top) and in the total duration of the simulation (bottom). The actual terrestrial planets are represented by icons in the background. The area marked in blue defines the semi-major axes range for the Venus and Earth analogs, represented by different symbols. Before this range are the Mercury analogs and after are the Mars analogs.}
  \label{CompVeMaMe}
\end{figure}

Figure \ref{CompVeMaMe} illustrates the delivery of comets on all of the terrestrial planet analogs from our successful simulations. The top panel shows that some terrestrial planets analogs have been supplied a lot of cometary material after the Earth analog has reached half of its mass, while others have not. This emphasises the stochastic nature of the cometary bombardment and explains how the terrestrial planets may have substantially different atmosphere compositions. Indeed, terrestrial planets might have accreted very variable amounts of cometary mass after having formed completely, thus disparately contributing to their atmospheres. 

Our results thus converge with those of \citet{Marty2016} who point the high abundance of $^{36}$Ar in the atmosphere of Venus compared to Earth, and suggest that the supply of atmospheric volatiles to the terrestrial planets most probably involved a small population of objects, resulting in an heterogeneous distribution of cometary isotopic signatures.

\citet{Avice2022} argue that the (late) contribution of comets on Venus was most probably more important than on Earth, given the much higher $^{36}$Ar/$^{22}$Ne ratio in Venus atmosphere (although the low precision on this measurement still allows for an Earth-like elemental ratio). Indeed, comets also have a significant $^{36}$Ar/$^{22}$Ne ratio, as they probably do not contain any Ne \cite{BarNun1998}.

Finally, it was proposed that the isotopic signatures of some volatiles in the Martian atmosphere (mainly N, Ar and Kr) are consistent with a cometary contribution \cite{Owen1995, Marty2016}. Our simulations outcomes support this hypothesis with 6.4 $\times$ $10^{18}$ to 5.5 $\times$ $10^{23}$ g of cometary mass being supplied to Mars analogs after the Earth analog has reached half of its mass (and Mars has either already finished or is about to finish its formation by this time).

Note that the bottom panel in Figure \ref{CompVeMaMe} displays the cumulative cometary mass that has been delivered to each of the terrestrial planets analogs during the 100 million years of the simulation, and these results have \textit{not} been \textit{normalized} to the moment of the instability (unlike the values in Table \ref{table2}). This part of the figure mainly shows that the cumulative cometary mass hitting the terrestrial planets is proportional to the final mass of these planets. Indeed, it is comprised between 2.2 $\times$ $10^{24}$ and 2.4 $\times$ $10^{25}$ g for Venus analogs (or between 2.6 $\times$ $10^{-4}$ and 2.7 $\times$ $10^{-3}$ of the Venus analogs' respective final masses); between 7.8 $\times$ $10^{23}$ and 5.0 $\times$ $10^{24}$ g for Earth analogs (or between 1.7 $\times$ $10^{-4}$ and 8.1 $\times$ $10^{-4}$ of the Earth analogs' respective final masses); and between 1.5 $\times$ $10^{23}$ and 8.5 $\times$ $10^{23}$ g for Mars analogs (or between 2.1 $\times$ $10^{-4}$ and 7.8 $\times$ $10^{-4}$ of Mars analogs' respective final masses). This is due to the fact that a bigger planet tends to accrete more embryos during its formation and the probability of collision between a comet and each of these constituent embryos adds up for the calculation of the total cometary contribution. However, if we take a closer look at the values of the cometary contributions relative to their respective terrestrial planets analogs' final masses, we find that Venus might have been supplied more comets than Earth or Mars. This result supports the claim made by \citet{Avice2022} mentioned above.

\subsection{Limitations}
\label{limitations}
The accuracy of our quantitative results for the range of potential cometary mass accreted by the Earth and the other terrestrial planets are limited by four main factors. 

First, the perfect merging approach used in most N-body simulations of terrestrial planet formation tends to underestimate the accretion timescale and overestimate the final masses of planets \cite{Emsenhuber2019, Burger2019, Emsenhuber2021, Haghighipour2022}. However, \citet{Walsh2016} and \citet{Deienno2019} have included collisional fragmentation in their simulations, and still obtained successful outcomes, starting from an initial total mass of $\simeq$ 2 - 2.5 $M_{Earth}$  for the inner ring of planetesimals. 

Second, all the assumptions for the typical noble gas isotopic composition in comets rely on the measurements made on one single comet, 67P/Churyumov-Gerasimenko. However, it was shown that volatile content, including noble gases content, in cometary ice reflects gas pressure and temperature under which they formed \cite{BarNun1985, Yokochi2012, Almayrac2022}. Consequently, if we postulate that all comets formed in the same region, that is in the same conditions of pressure and temperature, the noble gas content in 67P could then be representative of all comets \cite{Marty2016}, if the comets did not lose a significant portion of their volatile over their thermal histories \cite{Gkotsinas2022}. 

Third, we assume that there is no loss of volatile elements to space during collisions between comets and terrestrial embryos. In the case of the Earth, this assumption was shown to be roughly correct for collisions occuring after the Moon-forming impact \cite{Marty2016} but it becomes less valid in the case of giant impacts which likely involve significant atmospheric loss \cite{Genda2005}. Overall, models simulating the global volatile budget during giant collisions remain elusive.  \citet{deNiem2012} have shown that single impacts of very massive bodies are of more importance for the atmospheric evolution than other input parameters. In contrast, \citet{Sinclair2020} argue that only the smallest cometary impactors could contribute material to the atmosphere. In the case of the “dry” comets however, their density could be high enough that a non-zero fraction of even the largest objects could be accreted, given a low enough impact velocity.

Finally, we are limited by the number of particles in our N-body simulations, and the computational cost associated with it. Indeed, each simulation includes 10000 comets, having a mass of 1.5 $\times$ $10^{25}$ g ($\simeq$ Pluto-mass) each in order to have a total mass of 25 $M_{Earth}$ for the outer disk of planetesimals. It was shown that the combined mass of Pluto-mass objects in the initial outer disk of planetesimals could represent 10\% to 40\% of the total disk mass \cite{Nesvorny2016}. Therefore, our simulations mostly reflect those massive bodies carrying an important portion of the outer disk mass. The latter could be responsible for the stochastic component of the cometary bombardment on the planets of the inner solar system.
However, it would be more realistic to perform simulations also including a large number of smaller-sized comets, to account for the non-stochastic component of the bombardment.

Bearing all these considerations in mind, we believe that our simulation results, obtained from a dynamical model perspective, are sufficiently reliable as a first approximation given their consistency with geochemical constraints. In addition, both the perfect-merging approach and the limited number of comets have the particular advantage of reducing the overall computational cost of our simulations.

\section{Conclusion}
We have performed N-body simulations of the early solar system considering different set of initial conditions and instability cases. In our successful simulations, we have calculated the probability of collision between each pair of comet and terrestrial planet embryo through time and obtained an estimation of the cometary mass accreted by each of our Earth, Venus and Mars analogs as a function of time.

Given the fact that only a small number of collisions from big impactors may have been necessary to account for the cometary contribution to Earth, and that the primordial outer disk of planetesimals contained a non-negligible portion of Pluto-mass objects \cite{Nesvorny2016, Kaib2021}, we propose that the cometary bombardment on Earth included a stochastic component - dictated by Pluto-mass objects -, in addition to a smoother component - dictated by smaller comets. While the smoother component has been extensively investigated in \citet{Nesvorny2023}, we have focused on the stochastic component of the bombardment

First and foremost, we find that the contribution of comets on Earth might have been delayed with respect to the timing of the instability, due to this stochastic component. While the contribution of comets in the atmosphere of the Earth is mostly established \cite{Marty} and volatiles in the Earth's mantle were shown to be chondritic \cite{Peron, Broadley}, it remains unclear whether a few comets could have contributed to the deep mantle early in the Earth's accretion history \cite{Peron2021}. Given these geochemical constraints, the cometary delivery on Earth might have either happened \textit{both} during the earliest phases of Earth's growth and after its formation was finished, or \textit{after} - and only after - its formation was finished. These two cases have been observed as possible - yet not frequent - outcomes of our simulations. In particular, the case where comets are exclusively supplied after the Earth analog was formed only occurs in simulations which involve an extremely rapid formation of the Earth analog.  For this reason, we favour the other scenario and support the idea that comets contributed to the mantle's budget. It is also possible that early-delivered cometary noble gases were not retained efficiently in the mantle, and were thus sequestered in the atmosphere. This alternative theory will be investigated in future works.

In any case, our dynamical simulations demonstrate that the \textit{Early Instabilty} scenario and a late supply of cometary material on Earth are not mutually exclusive. As a direct consequence, the xenon constraints set by the terrestrial mantle and atmosphere are not necessarily in conflict with an instability happening during the first 10 Myr of the solar system. 

Above all, our study provides a broad overview of the possible cometary bombardment mass contribution and timings on Earth. More importantly, it highlights the stochastic variability of cometary delivery on the terrestrial planets.

\section{Acknowledgements}
Computer time for this study was partly provided by the computing facilities MCIA (Mésocentre de Calcul Intensif Aquitain) of the Université de Bordeaux and of the Université de Pau et des Pays de l'Adour, France. Numerical computations were also partly performed on the S-CAPAD/DANTE platform, IPGP, France. We acknowledge support from the CNRS MITI funding program (PhD grant to S.J.) R.D. was supported by the NASA Emerging Worlds program, grant 80NSSC21K0387. 

\medskip
\bibliography{reference}

\end{document}